%% file: ms.tex
\documentclass[iop]{emulateapj}

\input{macros}
\input{mle_params}

\begin{document}

\title{A Survey of \mgii Absorption at $2<z<6$ with Magellan /
  FIRE. \\ I: Sample and Evolution of the \mgii Frequency \altaffilmark{1}}
\shorttitle{Sample and Evolution of the \mgii Frequency}
\subjectheadings{Galaxies: evolution---Galaxies: halos---Galaxies: high-redshift---Infrared: general---intergalactic medium---quasars: absorption lines}
\author{Michael S. Matejek\altaffilmark{2}, Robert
  A. Simcoe\altaffilmark{2,3}} \altaffiltext{1}{This paper includes data gathered
  with the 6.5 meter Magellan Telescopes located at Las Campanas
  Observatory, Chile}\altaffiltext{2}{MIT-Kavli Center for
  Astrophysics and Space Research, Massachusetts Institute of Technology, 77 Massachusetts Ave., Cambridge, MA 02139, USA} \altaffiltext{3}{Sloan Foundation
  Research Fellow} 

\begin{abstract}
We present initial results from the first systematic survey for \mgii
quasar absorption lines at $z>2.5$.  Using infrared spectra of 46
high-redshift quasars, we discovered 111 \mgii systems over a path
covering $1.9<z<6.3$.  Five systems have $z>5$, with a maximum of
$z=5.33$---the most distant \mgii system now known.  The comoving \mgii
line density for weaker systems ($W_r<1.0$\AA) is statistically
consistent with no evolution from $z=0.4$ to $z=5.5$, while that for
stronger systems increases three-fold until $z\sim 3$ before declining
again towards higher redshifts.  The equivalent width distribution,
which fits an exponential, reflects this evolution by flattening as
$z\rightarrow 3$ before steepening again.  The rise and fall of the
strong absorbers suggests a connection to the star formation rate
density, as though they trace galactic outflows or other byproducts of
star formation.  The weaker systems' lack of evolution does not fit
within this interpretation, but may be reproduced by extrapolating low
redshift scaling relations between host galaxy luminosity and
absorbing halo radius to earlier epochs.  For the weak systems,
luminosity-scaled models match the evolution better than similar
models based on \mgii occupation of evolving CDM halo masses, which
greatly underpredict $dN/dz$ at early times unless the absorption
efficiency of small haloes is significantly larger at early times.
Taken together, these observations suggest that the general structure
of \mgii-bearing haloes was put into place early in the process of
galaxy assembly.  Except for a transient appearance of stronger
systems near the peak epoch of cosmic star formation, the basic
properties of \mgii absorbers have evolved fairly little even as the
(presumably) associated galaxy population grew substantially in
stellar mass and half light radius.
\end{abstract}

\section{Introduction}
\label{sec:intro}

\mgii absorption line searches provide a luminosity-independent method
for studying the distribution of gas in galactic haloes.  While this
technique has been employed for decades
\citep{weymann_mgii,lanzetta_mgii,tytler_mgii,sargentmgii,steidelandsargent,nestor2005,prochtersdss2006},
using tens of thousands of sightlines
\citep{prochtersdss2006,quiderSDSS}, these systematic surveys have all
focused exclusively on \mgii systems having $z<2.3$.  Only a handful
of \mgii absorption systems have been reported at redshifts $z>3$
based on serendipitous detections
\citep{elstonMgii4.38,kondoMgii3.5,jiangMgii4.8}.

In this paper we present a systematic survey of intergalactic \mgii
absorption at $2<z<6$ using QSO spectra taken with FIRE (the
Folded-port InfraRed Echellette) on the Magellan Baade Telescope.  By
characterizing the evolution of \mgii absorption at higher redshifts
we investigate the mechanisms which populate extended haloes with
magnesium during the epoch when the cosmic SFR approached and passed
its peak at $z\approx2$.

Currently, two theories have been advocated to explain the properties
of \mgii absorbers at low redshifts.  In one scenario, \mgii traces
cool clumps embedded in heated outflows from galaxies with high
specific star formation rates.  Alternatively, the distribution of
\mgii may reflect gravitational and gas accretion processes, as from
dynamical mergers or cold accretion filaments.

The outflow hypothesis is supported by direct observations of
blueshifted \mgii absorption in the spectra of star forming galaxies
\citep{ubiquitous2009}; \citet{rubin2010} have also observed this
trend and established a correlation between \mgii rest frame
equivalent width and star formation rate (SFR).  Moreover
\citet{bouche2006anticorrelation} cross-correlated $\sim 250,000$
luminous red galaxies with 1806 $W_r\gtrsim0.3\textrm{\AA}$ \mgii
absorbing systems at $z\sim 0.5$ from SDSS-DR3, finding an
anti-correlation between \mgii rest frame equivalent width and galaxy
halo mass.  This suggests that the individual \mgii systems are not
virialized, and the authors interpreted it as evidence that \mgii
systems trace star formation driven winds.

Concurrently, \citet{zibetti2007colors} showed with a sample of 2800
strong \mgii absorbers ($W_r>0.8\textrm{\AA}$) at low redshifts
($0.37<z<1.0$) that $W_r$ and galaxy color are correlated, with
stronger \mgii systems corresponding to the colors of blue
star-forming galaxies.  Follow up work
\citep{bouche2007ha,menardo2,noterdaeme2010,nestor2010strong} supports
this association of strong \mgii systems with star forming galaxies.

Yet not all \mgii absorbers are found around star forming galaxies.
In particular, galaxy-selected \mgii samples repeatedly show that star
forming and early-type systems alike give rise to halo absorption, in
proportions similar to the general field.  \citet{chen2010sfr} showed
that the extent of the \mgii halo increases with galaxy stellar mass
but correlates only weakly with specific SFR for a sample of 47 weaker
\mgii (mostly with $W_r<1\textrm{\AA}$) at $z<0.5$.  The authors
interpreted this as evidence that the \mgii absorbers reside in
infalling clouds that eventually fuel star formation.  In contrast to
Zibetti, \citet{chen2010lowz} see little correlation between $W_r$
strength and galaxy colors, a result corroborated by later research on
weak absorbers \citep{lovegrove2011,kacprzak2011incl}.
\citet{kacprzak2011incl} report a correlation between $W_r$ and galaxy
inclination in a sample of 40 \mgii absorption selected galaxies with
$0.3<z<1$ and $W_r\lesssim1\textrm{\AA}$.  This hints that some \mgii
absorbers may exhibit co-planar geometries and organized angular
momenta, as may be found in accreting streams, satellites, and
filaments, in contrast to star formation driven winds, which escape
perpendicular to the disk.

The different interpretations of the above studies may be understood
in terms of the selection processes used to construct each sample.
Broadly speaking, galaxy surveys around QSO sightlines pre-selected
for strong \mgii absorption tend to find systems with high specific
star formation rates
\citep{zibetti2007colors,bouche2007ha,menardo2,noterdaeme2010,nestor2010strong}.
Conversely, blind searches for \mgii absorption along QSO sightlines
serendipitously located near galaxies tend to find weaker \mgii
systems, and little correlation between \mgii incidence and SFR
\citep{chen2010sfr,chen2010lowz,lovegrove2011}.

Recently, the idea of a bimodal \mgii absorber population has gained
favor, with weaker absorbers $W_r\lesssim1\textrm{\AA}$ mostly tracing
distributed halo gas and stronger absorbers $W_r\gtrsim1\textrm{\AA}$
mostly tracing outflows \citep{kacprzak2011schech}.
This naturally explains the discrepant results between \mgiinsp-selected
galaxies and galaxy-selected \mgii absorbers: the two methods simply
sample different ends of the \mgii equivalent width distribution.
While this explanation is simple and appealing, the existence of
strong absorbers around fairly quiescent galaxies
\citep{gauthier2010,bowen2011,gauthier2011} suggests that in reality
there may be significant overlap between the populations in $W_r$
space.

The volume-averaged SFR, halo assembly and cold accretion rates, and
metal enrichment rate all vary with lookback time, but the restriction
of $z \lesssim 2$ for optical \mgii searches means that these surveys
do not sample epochs predating the SFR peak at $z\sim 2.5-3$.  This
inflection in the SFR density, along with the shutdown of cold
accretion at low redshift, could be a useful diagnostic for evaluating
the plausibility of the various mechanisms for distributing \mgii into
haloes.

While these factors motivate the study of \mgii systems at high
redshift, the endeavor has proven difficult in practice since the
observed-frame transition moves into the near-infrared.  In this
regime, absorption line searches are hampered by blending with OH sky
emission and telluric absorption, except at high resolutions where
instrument sensitivity becomes an issue.  We recently commissioned the
FIRE infrared spectrometer on Magellan with the aim of studying \civ
at $z\gtrsim 6$; its design characteristics also make it a powerful
survey instrument for mapping \mgii between $z\sim 2$ and $z\sim 5.5$
(and beyond as background targets are discovered).  This paper
describes the first results of an ongoing study of \mgii absorption in
early universe, based on an initial sample of 46 sightlines observed
with FIRE.

In Section \ref{sec:Data} we describe the data acquisition.  In
Section \ref{sec:analysis} we describe our analysis, including our
\mgii line identification procedure and completeness tests.  In
Section \ref{sec:res} we provide our main science results, including
binned and maximum likelihood fit population distributions, and
compare these results to those of previous studies with lower redshift
search ranges.  In Section \ref{sec:discuss}, we discuss the
implications of these results for the origin of \mgii absorbing
systems.  Throughout this paper we use a $\Lambda$CDM cosmology with
$\Omega_m=0.3$, $\Omega_\Lambda=0.7$, and $H_0=70$ km s$^{-1}$
Mpc$^{-1}$.

\section{Data Acquisition}
\label{sec:Data}

\begin{figure*}
\epsscale{1.0}
\plotone{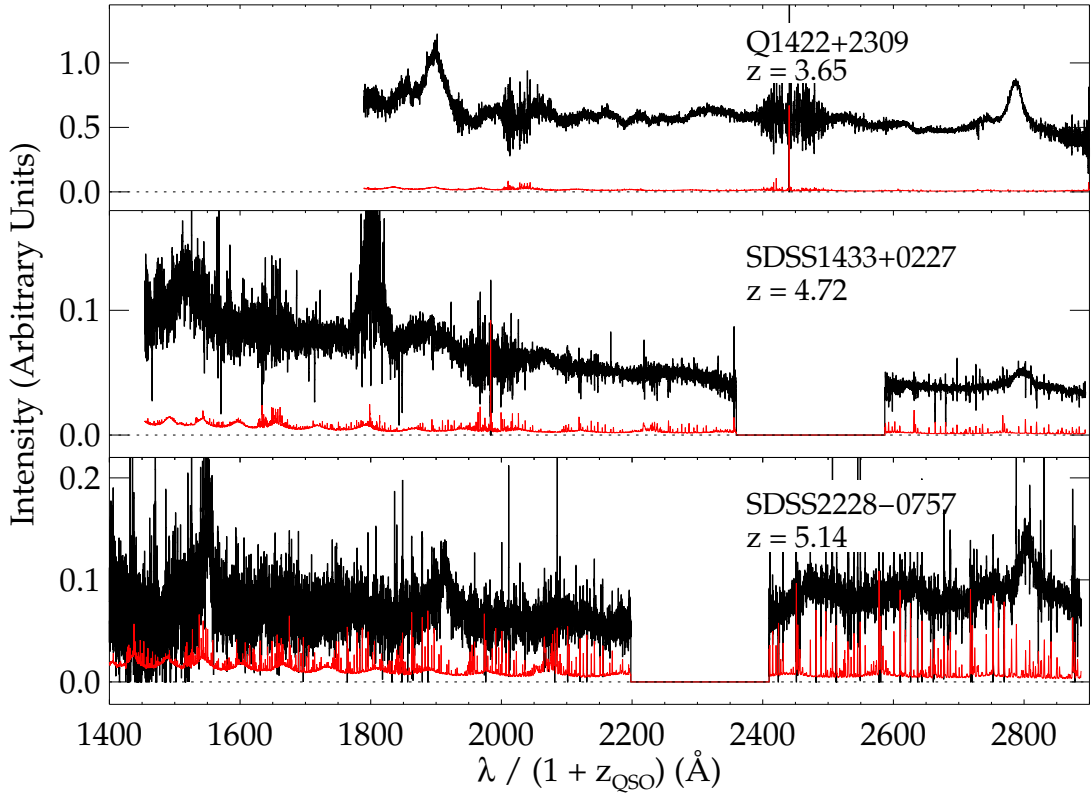}
\caption{Three example spectra arranged from top to bottom in order of
  decreasing signal-to-noise ratio (SNR).  The top spectrum is the
  highest-quality in the sample (SNR$=47$), the middle 
  is at the survey median quality (SNR$\approx 13$) and the bottom is one of the
  poorest quality spectra (SNR$=5$). The \mgii emission line is
  clearly visible at $\lambda_{rest}\approx 2800\textrm{\AA}$, as is a
  noisy region from poor telluric correction centered at $\sim1.14$
  $\mu$m in the observed frame ($z\sim3.1$ for \mgii absorbing
  systems.)}
\label{fig:spectra}
\end{figure*}

Our sample consists of 46 quasar spectra taken with FIRE
\citep{fireDesign2008,fireDesign2010}, between 2010 June and 
2011 April. This instrument is a single object, prism cross-dispersed
infrared spectrometer with a spectral resolution of R=6000, or
approximately 50 km/s, over the range 0.8 to 2.5 $\mu$m. We observed
using a 0.6'' slit and with a typical seeing of 0.5''-0.8''. The
wavelength coverage of FIRE set the lower redshift limit for our \mgii
absorption doublet search at $z\approx1.9$ in each sightline. The QSO
emission redshifts, which ranged from $z=3.55$ to $z=6.28$, typically
set the upper limits (although masked telluric regions described below
sometimes decreased these maximum cutoffs). Regions lying within
$3000$ \kms ~of each background QSO were excluded from the search
path, although the effect of this cut was negligible on our derived
results.  Table \ref{tab:qsoList} lists key observational properties
of the 46 quasars.  The quasars themselves were predominantly chosen
from the SDSS DR7 quasar catalog \citep{schneider}, with an
additional, somewhat heterogeneous selection of bright, well known
objects from the literature.

\input{qsoList}

We reduced the data using a custom-developed IDL pipeline named FIREHOSE,
which has been released to the community as part of the FIRE instrument
package. This pipeline evolved from the optical echelle reduction
software package MASE \citep{mase}. FIREHOSE contains many
important modifications for IR spectroscopy, including using terrestrial
OH lines imprinted on science spectra for wavelength calibration.

To correct for telluric absorption features, we obtained spectra of
A0V stars at comparable observing times, airmasses and sky positions
to our observed QSOs. We then performed telluric corrections and
relative flux calibrations using the xtellcor software package
released with the spextool pipeline \citep{cushing2004spextool}, which
employs the method of \citet{vacca2003tell}.  Regions lying between
the $J$ and $H$ bands (1.35 $\mu$m to 1.48 $\mu$m, excluding $z=$3.8
to 4.3 for \mgii systems) and the $H$ and $K$ bands (1.79 $\mu$m to
1.93 $\mu$m, excluding $z=$5.4 to 5.9), were masked out of the search
path. Another smaller region of path length $\Delta z\sim0.2$ centered
on $z\sim3.1$ (near the $Y$-$J$ transition) showed poor telluric
residuals (noticeable in Figure \ref{fig:spectra}).  We left this
region in our \mgii search since it was still possible to find strong
systems (we found six).  We report these systems in our line lists
but exclude this path and associated systems from our population
statistics as the sample completeness is low and variable from one
sightline to the next.

\begin{figure}
\epsscale{1.0}
\plotone{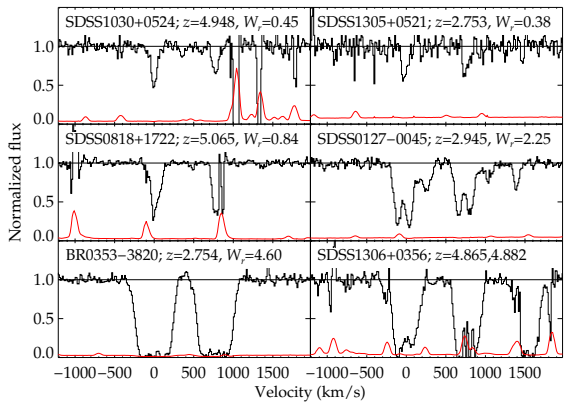}
\caption{Representative examples of \mgii absorption from the survey.
  These are chosen to illustrate some of the common features in the
  infrared data, including blending with sky emission lines, and
  variable signal-to-noise ratios.  Several very strong, and saturated
  \mgii systems are present in the data, sometimes with complex
  velocity structure.  In the bottom right panel, we see a pair of
  \mgii doublets with blended $2796/2803$ components, separated by
  $750$ km/s.}
\label{fig:sample_doublets}
\end{figure}

The final signal-to-noise ratios per pixel for our full QSO set ranged
from 4.0 to 47.2, with a median value of 12.9.  Figure
\ref{fig:spectra} shows the \mgii search portions for our highest
(top panel), a typical (middle panel), and a relatively poor (bottom panel)
signal-to-noise spectra.

\section{Analysis}
\label{sec:analysis}

\subsection{\mgii Line Identification}
\label{sub:mgii_search}

\subsubsection{Continuum Fitting}
\label{sub:cont_fit}

With just 46 QSO sightlines in our sample, it would be reasonable to
fit each continuum manually.  However, as part of our downstream
analysis we also run extensive Monte Carlo tests to characterize
sample completeness.  We wished to capture the effects of the
continuum fitting on our completeness.  Accordingly, we developed an
automated algorithm for continuum fitting, to make treatment of the
Monte Carlo sample both tractable, and as consistent as possible with
our handling of the true data.

First, we generated an initial line mask to prevent absorption and
emission features (including miscorrected skylines) from biasing the
continuum.  Next, we performed iterative, sigma-clipped linear fits on
the masked data over small segments of the spectra ($\sim$100 pixels,
or $\sim$1250 km/s, wide) in order to determine a list of spline knots
(with two knots drawn from each of these small segments). We then
implemented a cubic spline fit through these knots to produce a
continuum fit. Since knowledge of the continuum fit facilitates the
creation of the initial line mask, we iterated this between two and
five times before converging on a final continuum fit.

\subsubsection{Matched Filter Search}
\label{sub:mf}

After obtaining continuum-normalized spectra, we ran a matched filter
search with a signal-to-noise ratio cut of 5 to determine an initial
candidate list. The filter kernel consists of a pair of Gaussian
profiles separated by the \mgii spacing with FWHM ranging from 37.5 to
150 km/s. Sky subtraction residuals near OH lines are a source a noise
in the infrared, so we identified and masked badly miscorrected sky lines
before running this convolution to reduce the number of false
positives. Even with this masking, this matched filter search
typically led to $\sim$50 to 100 candidates per QSO sightline, of
which only $\sim2$ per sightline were true detections.

To mitigate the high false positive rate of the matched filter
procedure, we fit each candidate doublet with a pair of Gaussian
absorption profiles and subjected the resulting fit parameters to a
set of consistency checks (e.g. $W_{2796}\ge W_{2803 }$ within
errors). This additional doublet fitting filter step reduced the
number of false positives by a factor of $\sim$10, and left us with
256 candidates across all QSO sightlines.

\subsubsection{Visual Inspection}
\label{sub:inter}

We subjected these machine generated candidates to a visual inspection
to produce the final \mgii sample list. This interactive step lowered
the final list to 110 \mgii systems (plus one proximate system). Despite our best efforts to
automate the process, we found that the poorly telluric corrected
regions and abundance of sky lines mandated this visual inspection.

We also visually combed through all 46 spectra to identify systems
missed by the automated line finder. We found three systems with this
visual search that our automated line finding algorithm rejected.  One
system was a large 2.62\AA ~system that was ``too complex'' (and therefore was fit poorly by a
Gaussian absorption profile and consequently rejected), and the other
two were smaller systems (confirmed by FeII lines) that were partially obscured by sky
lines. The two smaller systems were manually added to our line list in Table
\ref{tab:dlist}, but were not included in statistical
calculations so as not to bias our results by counting them and then
overcorrecting for completeness.  The large system's equivalent width was greater than the highest value ($W_r=0.95\textrm{\AA}$) probed in our Monte Carlo completeness tests.  We are $\sim 100\%$ complete to such large systems, so this system was included in our statistical calculations.

\subsubsection{Final \mgii Sample}
\label{sub:measure}

\begin{figure*}
\epsscale{1.1}
\vskip 0.2in
\plotone{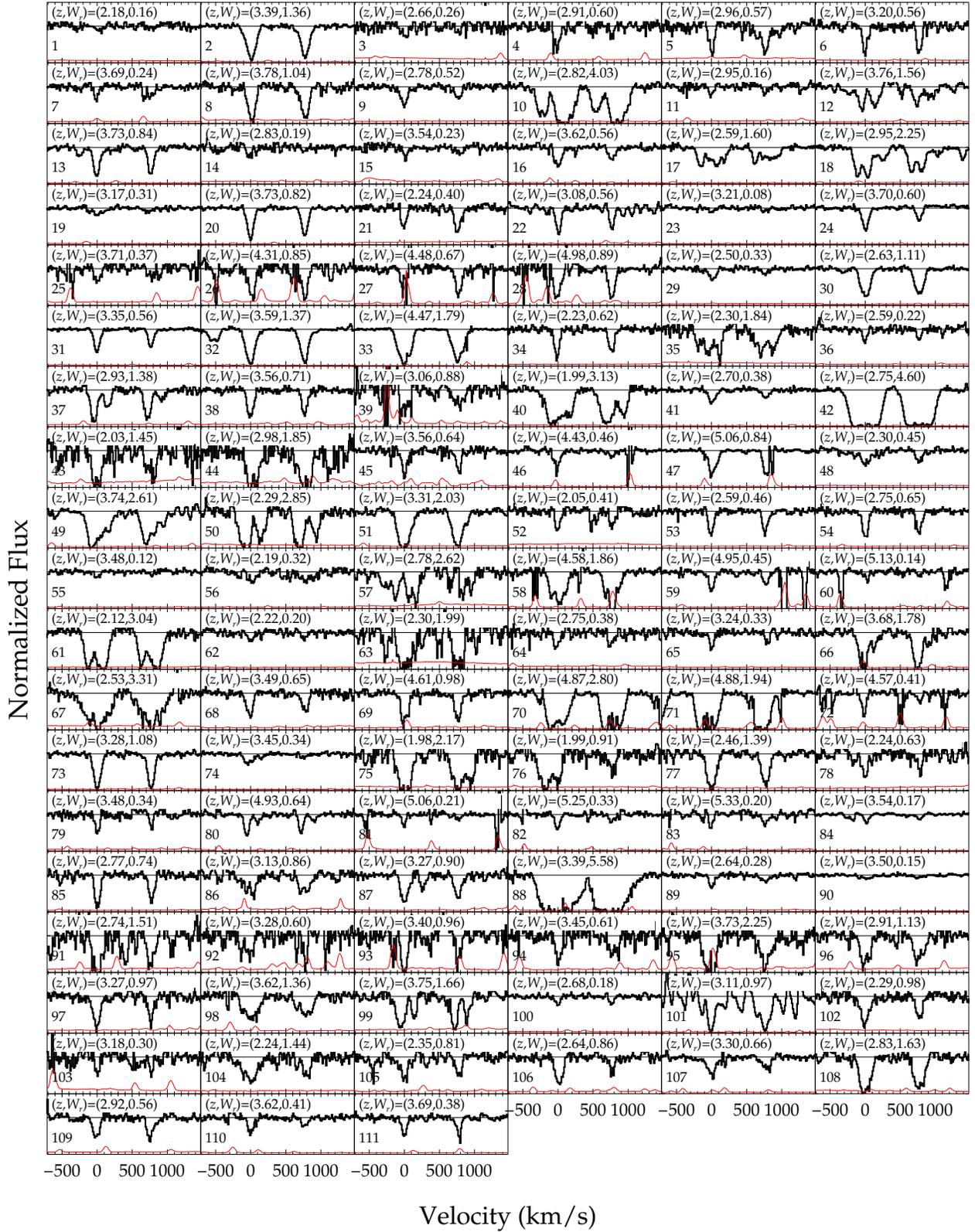}
\caption{The full FIRE \mgii sample, in continuum-normalized units,
  overplotted with errors (dotted line). The bottom edge of each plot
  corresponds to 0, and the horizontal line near the top to 1. Panel
  numbers correspond to entries in Table \ref{tab:dlist}.}
\label{fig:all_doublets}
\end{figure*}

\input{dlist}
\setcounter{table}{1}
\input{dlistb}

The final \mgii line list, as well as the main properties of each
system, are listed in Table \ref{tab:dlist}.  Figure
\ref{fig:sample_doublets} shows a selection of six \mgii systems in
detail, while Figure \ref{fig:all_doublets} shows continuum normalized
postage stamps of the entire sample.

\subsection{Simulations of Completeness and Contamination}
\label{sub:sims}

The nature of the infrared sky suggests that our survey completeness
will vary strongly with redshift as systems overlap with atmospheric
OH features.  The common method of calculating sensitivity from 1D
error vectors, usually estimated assuming photon statistics of the
background, may not capture sky subtraction residuals completely.  For
this reason we have performed extensive simulations to characterize
completeness and contamination in both the automated and interactive
portions of our line identification procedure.

\subsubsection{Automated Completeness Tests}
\label{sub:mc_auto}

To test the automated portion of our identification pipeline, we ran a
large Monte Carlo simulation in which we injected artificial \mgii
systems into our QSO spectra and then ran our line identification
algorithm to determine success/recovery rates.  The full Monte Carlo
completeness simulation involved 20,000 simulated spectra per QSO
injected with \mgii at an inflated rate of $dN/dz\approx5$, which
combined led to over $12$ million injected systems. The velocity
spreads of the injected systems were calibrated to mimic the behavior
of true, detected systems (see Figure \ref{fig:vspreads}, below). The
injected systems were distributed uniformly in rest frame equivalent
between $0.05\textrm{\AA}<W<0.95\textrm{\AA}$.  (A separate test
showed that the completeness did not improve significantly beyond this
upper limit.) Our small number of QSOs allowed us to calculate
individual completeness matrices for each sightline $q$, in finely
tuned redshift ($dz=0.02$) and rest frame equivalent width
($dW=0.01\textrm{\AA}$) bins. We denote these values, which reflect
the completeness of the automated line search alone, as $L_{q}(z,W).$

\subsubsection{False Positive and Spurious Rejection Tests}
\label{sub:mc_inter}

Because our identification algorithm involves an interactive
evaluation of each system, the user may reject true \mgii systems or
accept spurious doublet candidates caused by miscorrected skylines or
correlated noise. We developed a simple test to quantify these errors
using simulated data. First, we created a large number of simulated
spectra in which we injected \mgii doublets using a process identical
to that described in Section \ref{sub:mc_auto} above, but with a
slightly larger doublet velocity spacing. We ran these artificial
spectra through our identification and sample cleaning algorithms, and
interactively accepted/rejected over 1500 such candidates.  By using
nonphysically spaced doublets, we were guaranteed that auto-identified
candidates were either correlated noise or simulated \mgii doublets.
The human decision accuracy for each of these cases was translated
into success rates, which were then folded into our completeness
results.

\begin{figure}
\epsscale{1.0}
\plotone{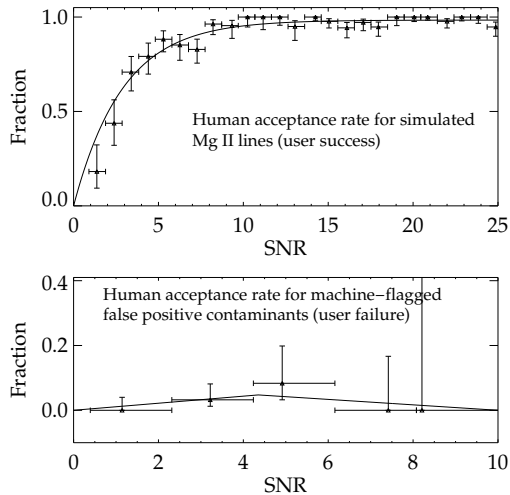}
\caption{Calibration of user accuracy for interactive acceptance of
  \mgii doublets and rejections of false-positive candidates.  The
  individual points represent the binned rates, and the overplotted
  solid lines are the maximum likelihood fits to the functional forms
  in Equations \ref{eq:p_mgii} (top panel) and \ref{eq:p_fp} (bottom
  panel). Displayed errors bars on the binned acceptance rates
  represent the Wilson score interval.}
\label{fig:user_rates}
\end{figure}

The time consuming nature of this interactive test did not allow for
the fine grained calculations produced above in the automated
completeness tests. We found, however, that our ability to distinguish
both false positives and true systems scaled strictly with the
signal-to-noise ratio of the detected candidates.  We therefore used
this simulation to parametrize these success rates as a function of
line SNR, and applied the resulting scalings to the larger Monte Carlo
completeness simulation and true data sets.

Specifically, for each candidate we calculate a boxcar SNR $s\equiv
W_r/\sigma_{W_r}$ of the 2796 \AA ~line. We then tabulated the
user-accept percentages of both the injected systems ($P^{\mmgii}(s)$;
ideally 100\%), and the false positives ($P^{\rm FP}(s)$, ideally
0\%) as a function of the SNR by calculating maximum
likelihood fits to chosen functional forms.  For the injected systems,
we chose an exponential for $P^{\mmgii}$,
\begin{equation}
P^{\mmgii}(s)\equiv P_{\infty}\left(1-e^{-s/s_{c}}\right),\label{eq:p_mgii}
\end{equation}
where $P_{\infty}$ (MLE estimate: $0.984$) is the probability that the
user accepts a true system with large SNR and $s_{c}$ (MLE estimate:
$2.89$) is an exponential scale factor.  Note that even at large SNR
the acceptance rate is not 100\% because of residual regions with very
bright sky lines and/or strong telluric absorption.  The binned
acceptance rates of injected systems, as displayed in the top panel of
Figure \ref{fig:user_rates}, motivated this functional form for
$P^{\mmgii}(s)$.

To estimate the user acceptance rate of contaminating false positives,
we chose a triangle function for $P^{\rm FP}$,
\begin{equation}
P^{\rm FP}(s)\equiv \left\{
\begin{array}{ll}
P_{\rm max}^{\rm FP}\left(\frac{s}{s_{p}}\right) & \hspace{1em}s\le s_{p} \\
P_{\rm max}^{\rm FP}\left(\frac{s-s_{f}}{s_{p}-s_{f}}\right) & \hspace{1em}s>s_{p}
\end{array}\right.
\label{eq:p_fp}
\end{equation}
where $P_{\rm max}^{\rm FP}$ (MLE estimate: $.0474$) is the maximum
contamination rate, which occurs at a SNR of $s_{p}=4.35$.  By SNR
$s_{f}\equiv 10$, the user-error returns to zero again. The property
that the false positive acceptance rate approach zero at both $s=0$
(the user never accepts a system with low SNR) and large
signal-to-noise (the user rarely makes mistakes at high SNR) in
conjunction with the lack of good motivation for a more complicated
form led to the functional form in Equation \ref{eq:p_fp}. The bottom
panel in Figure \ref{fig:user_rates} displays these binned
acceptance rates and the overplotted maximum likelihood fits.  Since
$P^{\mmgii}$ and $P_{\rm FP}$ are estimates of the success rate $p$ of
a binomial distribution, we use the Wilson Score interval to determine
the uncertainty in $p$, shown as error bars in each bin.  The large
errors in the bottom panel at high SNR reflect a paucity of high-SNR
false positives in the sample.  While the formal uncertainty in $p$
is large, such strong false positives are rare and were correctly
identified in the few cases where they arose.

For each individual sightline, the total acceptance rates $P^{\mmgii}$
and $P^{\rm FP}$ were estimated as a function of $W_r$ using the
continuum-normalized error arrays to calculate the SNR.  We denote the
resulting grids of user-acceptance rates as $A^{\mmgii}(z,W)$ and
$A^{\rm FP}(z,W)$.  The total completeness $C_{q}(z,W)$ for each QSO
$q$ is then the product of the fraction of systems identified by the
line finder $L_{q}(z,W)$ and the fraction of systems passing visual
inspection,
\begin{equation}
C_{q}(z,W)=L_{q}(z,W)A_{q}^{\mmgii}(z,W).\label{eq:comp_tot_q}
\end{equation}

\subsubsection{Survey Completeness}
\label{sub:comp_results}

\begin{figure}
\epsscale{1.1}
\plotone{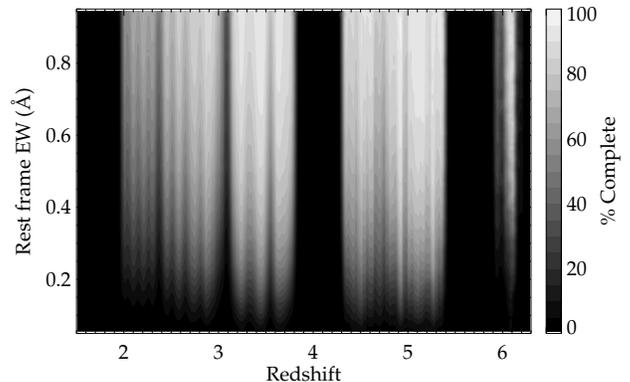}
\caption{The pathlength-weighted average completeness $C(z,W)$ as a
  function of redshift $z$ and equivalent width $W_{r}$, for all
  survey sightlines combined. The error in this estimate is typically
  $\lesssim5\%$ within our the search range.}
\label{fig:complete_image}
\end{figure}

\begin{figure}
\epsscale{1.0}
\plotone{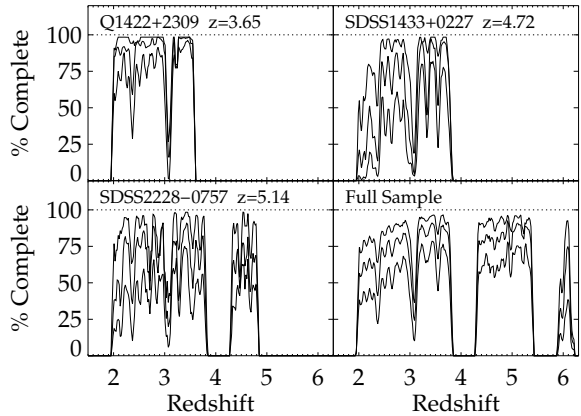}
\vskip 0.15in
\caption{Sightline-specific completeness for three fiducial equivalent
  widths $W_r$: 0.3, 0.5, and 1.0$\textrm{\AA}$ (from bottom to top in
  each panel). The first three panels correspond to the three QSO spectra shown in Figure
  \ref{fig:spectra}, and typify the best, median, and a poor SNR in our
  sample, respectively. The fourth panel is the average across the
  full survey.}
\label{fig:complete_slices}
\end{figure}

The total pathlength-weighted completeness across all sightlines
is shown in Figure \ref{fig:complete_image}. Figure
\ref{fig:complete_slices} shows completeness calculations at three
fiducial values of $W$ across the full survey and also for the three
QSO spectra depicted in Figure \ref{fig:spectra}. The acceptance rates
for false positives do not enter our simulated completeness results
here, but will be combined with the rejected candidate distribution
later in Section \ref{sub:adj_fp} in order to adjust the population
distribution directly.

We can use our completeness calculations for each QSO $C_{q}(z,W)$ to calculate
the redshift path density $g(z,W)$ of our survey, defined as the
total number of sightlines at a given redshift $z$ for which \mgii
systems with rest equivalent widths greater than $W$ are observable.
Typically, this quantity is defined with a hard cutoff \citep{steidelandsargent,nestor2005,prochtersdss2006};
mathematically, 
\begin{equation}
g(z,W)\equiv\sum_{q=1}^{Q}R_{q}(z)\theta(W-W_{q}^{\hbox{min}}(z)),\hspace{1mm}\textrm{hard cutoff}\label{eq:gz_old}
\end{equation}
where the sum is over QSOs, $Q$ is the total number of sightlines
probed, $\theta(x)$ is the Heaviside function, and $W_{q}^{\hbox{min}}(z)$
is the rest equivalent width threshold floor. The factor $R_{q}(z)$
represents the \mgii search region probed for QSO $q$, and is 1 for
every value of $z$ for which a \mgii system of any size could conceivably
have been found, and 0 everywhere else. (More specifically, it is
1 at every redshift within the minimum and maximum redshift \mgii search
limits for QSO $q$, except for those regions excluded because of
poor telluric corrections.) The threshold floor $W_{q}^{\hbox{min}}(z)$
here sets the limit on discovered \mgii systems that are included in
the final analysis: \mgii systems located at redshift $z_{m}$ for
QSO $q$ with rest equivalent widths $W_{m}>W_{q}^{\hbox{min}}(z_{m})$ are
included, while smaller absorption systems are not. These threshold
floors are specifically chosen such that the QSO spectra have (at
least roughly) 100\% completeness above these levels.

If completeness is explicitly quantified, however, then it may be
folded directly into the pathlength calculations. For example, having
two QSO sightlines that are 50\% complete at a given redshift is as
good as having one which is 100\% complete. If known, this completeness
may be folded into our expression for the redshift path density, 
\begin{eqnarray}
g(z,W) & \equiv & \sum_{q=1}^{Q}R_{q}(z)C_{q}(z,W),\hspace{1mm}\textrm{general}\label{eq:gz_new}
\end{eqnarray}
where $C_q(z,W)$ is the user-error adjusted completeness from Equation \ref{eq:comp_tot_q}.
Comparing this equation with Equation \ref{eq:gz_old} above for
the redshift path density with a hard threshold floor reveals that
such a floor amounts to approximating the completeness as either
0 or 1 at all redshifts and rest frame equivalent widths,
\begin{equation}
C_{q}(z,W)=\left\{
\begin{array}{ll}
1 & \hspace{1em}W>W_{q}^{\hbox{min}}(z),\hspace{1mm}R_q(z)=1\\
0 & \hspace{1em}\textrm{otherwise}
\end{array}\right.
\label{eq:comp_floor}
\end{equation}
Unlike Equation \ref{eq:gz_old} for $g(z,W)$ above, this formulation
in Equation \ref{eq:gz_new} does not simply exclude regions of
the data which, although perhaps not 100\% complete, still contain
valuable information.

\begin{figure}
\epsscale{1.0}
\plotone{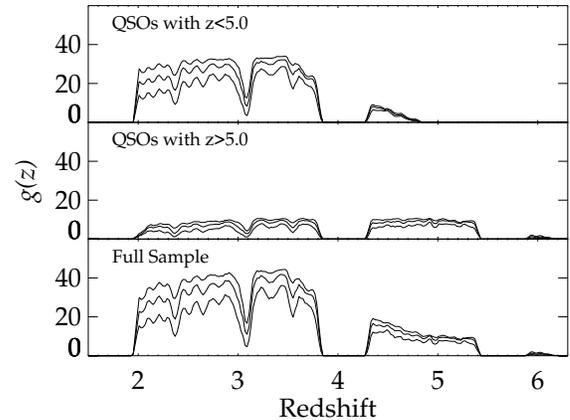}
\vskip 0.15in
\caption{The completeness-weighted number of sightlines over which we
  are sensitive at three specific rest equivalent widths: $0.3,
  0.5$, and 1.0$\textrm{\AA}$ (from bottom to top in each panel). The
  top panel corresponds to the 35 quasars in our survey with redshift
  less than 5, the middle panel to the 11 quasars with redshift
  greater than 5, and the bottom panel to the full survey.}
\label{fig:gz}
\end{figure}

\begin{figure}
\epsscale{1.0}
\plotone{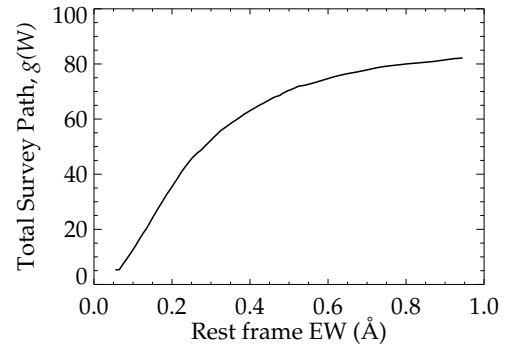}
\caption{The total path $g(W)$ as a function of rest frame
equivalent width, as given in Equation \ref{eq:tot_path}.}
\label{fig:gw}
\end{figure}

The redshift path densities $g(z,W)$ for the QSOs in our survey with
$z<5$, $z>5$, and all $z$ are shown in the three panels of Figure
\ref{fig:gz}. Figure \ref{fig:gw} displays the total path $g(W)$
as a function of rest frame equivalent width, found by integrating
the redshift path density $g(z,W)$ over the full redshift search
range of our survey,
\begin{equation}
g(W)\equiv\int g(z,W)dz.\label{eq:tot_path}
\end{equation}

\subsubsection{Parameter Errors}
\label{sub:errors}

\begin{figure}
\plotone{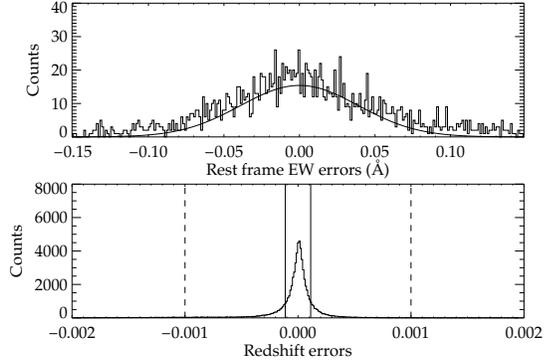}
\caption{Characterization of measurement errors from our Monte Carlo
  simulations.  The top panel shows equivalent width errors calculated
  by comparing measured to injected values; we found zero mean offset
  and $\sigma_{W}=0.055\textrm{\AA}$.  The bottom panel shows errors
  in measured redshift, with zero mean offset and
  $\sigma_{z}=2.74\times 10^{-4}$.  For reference, this error is
  comparable to the size of one FIRE pixel
  ($dz^{\textrm{pix}}=2.25\times 10^{-4}$), which is indicated with
  solid vertical lines.  The dotted vertical lines represent the
  precision provided in Table \ref{tab:dlist}.}
\label{fig:sim_errors}
\end{figure}

We also used our Monte Carlo simulation to study the errors in our
measured line parameters. Figure \ref{fig:sim_errors} shows
histograms of the differences in the injected and measured rest frame
equivalent widths (top panel) and redshifts (bottom panel).  The rest
frame equivalent width plot contains the errors for over 1750 \mgii
$2796\textrm{\AA}$ and $2803\textrm{\AA}$ singlets fit by hand. The
overplotted zero mean Gaussian has a width equal to the standard
deviation of the sample, $\sigma_{W}=0.055\textrm{\AA}$.  The plot
shows no significant biases.  The redshift plot contains errors for
the first 200 simulated spectra for each QSO in our Monte Carlo
completeness simulation (corresponding to over 67,500 systems). The
solid vertical lines are separated by one pixel
($dz^{\textrm{pix}}=2.25\textrm{e}-4$), and the dotted vertical lines
represent the precision given in Table \ref{tab:dlist}.  The
standard deviation of the sample is $\sigma_{z}=2.74\textrm{e}-4$.

\section{Results}
\label{sec:res}

\subsection{Population Statistics}
\label{sub:table_stats}

\begin{figure}
\epsscale{1.0}
\plotone{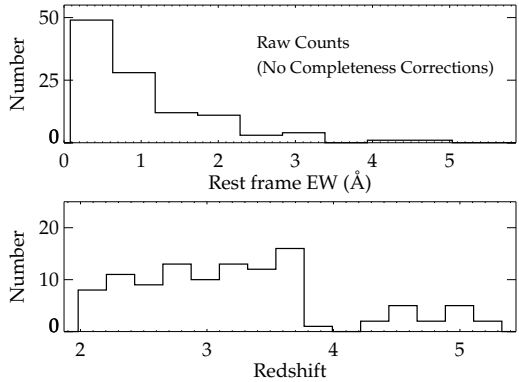}
\caption{Histograms over rest frame equivalent width (top panel) and
  redshift (bottom panel) for the 110 \mgii systems found in our
  survey and listed in Table \ref{tab:dlist} (proximate system not
  included). These plots represent raw counts; they do not include
  corrections for incompleteness or false-positive contamination.}
\label{fig:sample_hists}
\end{figure}

\begin{figure}
\epsscale{1.0}
\plotone{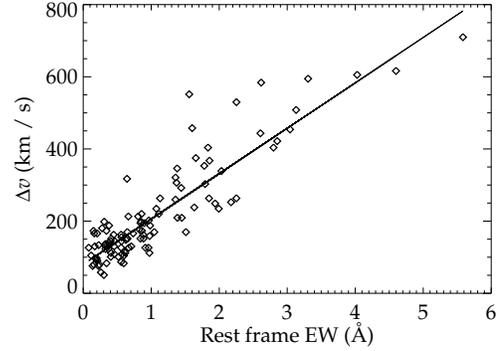}
\caption{The velocity spreads (in km/s) plotted against the
  2796$\textrm{\AA}$ singlet rest frame equivalent width for the 110
  located systems listed in Table \ref{tab:dlist} (proximate system
  not included). The overplotted line is the sigma-clipped best fit,
  $\Delta v=(125.7\pm6.0\textrm{
    km/s/\AA})W_{r}(2796)+(80.2\pm8.6\textrm{ km/s})$.}
\label{fig:vspreads}
\end{figure}

Figure \ref{fig:sample_hists} shows raw histograms of the rest frame
equivalent width (top panel) and redshift (bottom panel) distribution
of the sample's 110 \mgii systems, before corrections for incompleteness (proximate system not included).
The rest frame equivalent widths range from $W_r=0.08\pm0.01\textrm{\AA}$
(SDSS0140-0839, $z=3.71$) to $W_r=5.58\pm0.07\textrm{\AA}$
(CFQS1509, $z=3.39$). The sample contains $5$ \mgii systems with
redshifts over 5, with a maximum value of $z=5.33$ (SDSS1411, $W_r=0.20\textrm{\AA}$).
Three of these systems come from SDSS1411's sightline, which contains
four high redshift systems at $z=4.93,$ 5.06, 5.25, and 5.33.

Figure \ref{fig:vspreads} displays the relationship between the rest
frame equivalent widths of the $2796\textrm{\AA}$ \mgii singlets and
velocity spreads (measured to where the absorption troughs intersect
the continuum) of this sample. These velocity spreads have been
adjusted for FIRE's resolution of 50 km/s. The overplotted line is the
sigma-clipped best fit line, given by
\begin{eqnarray}
\Delta v=(125.7\pm6.0)\left({{W_{r}}\over{1 {\rm \AA}}}\right)+(80.2\pm8.6) {\rm ~km ~s}^{-1}. \label{eq:dv}
\end{eqnarray}
This slope and the strong correlation present in Figure
\ref{fig:vspreads} are consistent with previous results from
\citet{ellisonEW} at lower redshifts, $0.2\lesssim z \lesssim2.4$. We
incorporated this measured correlation and scatter into our Monte
Carlo completeness simulations described in \ref{sub:mc_auto} to
ensure that our injected systems resembled true systems from the
sample. (Strong injected systems were modeled as a blend of
  narrow components with total velocity spread drawn from the
  distribution specified by Equation \ref{eq:dv}).

\subsection{Incorporating Variable Completeness in Population Distributions}
\label{sub:dndz_math}

When calculating the line density and equivalent width distributions,
we first define the true number of systems in the universe within a
binned region $k$ centered on $(z_k,W_k)$ as $N_{k}$.  If the total
redshift path in bin $k$ is $\Delta z_{k}$, and the total range of
rest frame equivalent width in the bin is $\Delta W_{k}$, then the
population densities are given by
\begin{eqnarray}
\left(\frac{d^{2}N}{dzdW}\right)_{k} & = & \frac{N_{k}}{\Delta z_{k}\Delta W_{k}}\label{eq:pops_d2ndzdw}\\
\left(\frac{dN}{dz}\right)_{k} & = & \frac{N_{k}}{\Delta z_{k}}\label{eq:pops_dndz}\\
\left(\frac{dN}{dW}\right)_{k} & = & \frac{N_{k}}{\Delta W_{k}}.\label{eq:pops_dndw}
\end{eqnarray}

Using our completeness matrices we may relate these quantities to the
expected number of survey detections. A complete derivation of this
method is given in the Appendix. In summary, for an incomplete survey
with path length given by $g(z,W)$, the number of detected systems in
a bin $k$ is estimated as
\begin{equation}
\breve{N}_{k}(z,W)=\int\int_{k}g(z,W)\frac{d^{2}N}{dzdW}dzdW.\label{eq:n_est}
\end{equation}
The average completeness $\overline{C}_{k}$ is defined as the ratio
between this and the true number of systems $N_{k}$,
\begin{eqnarray}
\overline{C}_{k} & \equiv & \breve{N}_{k}/N_{k}\\
& = & \frac{\int\int_{k}\left(\sum_{q=1}^{Q}R_{q}(z)C_{q}(z,W)\right)\frac{d^{2}N}{dzdW}dzdW}{\int\int_{k}\left(\sum_{q=1}^{Q}R_{q}(z)\right)\frac{d^{2}N}{dzdW}dzdW}.\label{eq:cdef_main}
\end{eqnarray}

Our Monte Carlo calculations give us access to $\overline{C}_{k}$
(which in turn leads to $N_{k}$) with an important caveat. The average
completeness within a bin is a weighted integral
of the fine grained completeness across the bin, with weighting
function $d^{2}N/dzdW$.  If the true completeness varies within a bin,
then that average depends on the very function we are trying to
calculate.

There are three ways around this dilemma: (1) We can assume that the
completeness $C_{q}(z,W)$ is constant across all sightlines and
across the full breadth of the bin. Since the quality of the data
varies largely from spectrum to spectrum, this assumption is not
warranted. (2) We can assume that the population density within the
bin is approximately constant. With this approximation, the
population density in bin $k$ reduces to
\begin{equation}
\left(\frac{d^{2}N}{dzdW}\right)_{k}=\frac{\breve{N}_{k}}{\int_{k}g(z,W)dzdW}.\label{eq:pop_dens_approx}
\end{equation}
This is the method employed in the recent \civ study of
\citet{simcoeCiv} when considering variable completeness within bins,
and implicitly in any previous \mgii studies which employed the hard
threshold completeness floor given previously in Equation
\ref{eq:comp_floor}.  (3) We can compensate for this variability in
both completeness and absorber line density by employing a maximum likelihood estimate to a functional form
of the frequency distribution in the calculation of the average
completeness in Equation \ref{eq:cdef_main}.

We explored methods (2) and (3) using the maximum-likelihood estimates of
the frequency distribution defined later in this paper.  Figure
\ref{fig:study_dndz_effects} shows the effect of these corrections on
$dN/dz$ for the range $W>0.3\textrm{\AA}$.
``X''-shaped points indicate our $dN/dz$ values prior to completeness
correction; the triangles show completeness-corrected values with no
user-screening (i.e., using $\overline{L}$ rather than $\overline{C}$
for the correction and not adjusting for false positives, as described next section).  The squares and diamonds include the full
completeness and false positive corrections using methods (2) and (3) respectively to
treat intra-bin variation in the frequency distribution.  Apparently
this effect contributes a negligible correction to our calculations of
the population densities when compared with the sample's Poisson
errors. For this reason, we will instead use method (2) for
calculating $\overline{C}$ unless otherwise stated.

\subsection{Adjusting for False Positives}
\label{sub:adj_fp}

In the previous section, we calculated population densities in the
presence of variable completeness, but did not account for false
positives. Adjustments for false positives once again require Equations
\ref{eq:pops_d2ndzdw} to \ref{eq:pops_dndw} for the population
densities, but with the true number of systems $N_{k}$ calculated as
\begin{equation} \label{eq:nfp}
N_{k}=\frac{\breve{N}_{k}(1-\overline{A}_{k}^{F})-\overline{A}_{k}^{F}\breve{F}_{k}}{\overline{C}_{k}-\overline{L}_{k}\overline{A}_{k}^{F}},
\end{equation}
where $\breve{N}_{k}$ is the number of detected \mgii systems in bin
$k$, $\breve{F}_{k}$ is the number of rejected candidates,
$\overline{C}_{k}$ is the average completeness, $\overline{L}_{k}$ is
the automated line identification finding probability, and
$\overline{A}_{k}^{F}$ is the user acceptance rate of non-\mgii
candidates. All average values here are path length weighted (once
again employing the assumption that the population distributions are
approximately constant across the bin). A full derivation of this
formula is given in the Appendix. Note that as
$\overline{A}_{k}^{F}\rightarrow0$ (the user correctly rejects all
false positives), this reduces to the previous formula,
$\overline{C}_{k}=\breve{N}_{k}/N_{k}$.

\begin{figure}
\epsscale{1.1}
\plotone{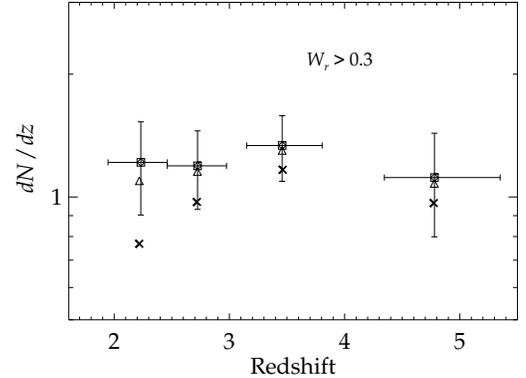}
\caption{Illustration of the effect of incompleteness corrections for
  systems with $W>0.3\textrm{\AA}$.  ``X"-shaped points show the line
  density prior to completeness correction.  Triangles incorporate
  the automated completeness $L_{q}(z,W)$ but no adjustment for
  interactive rejection of candidates.  Squares further incorporate
  the additional adjustment for the user-rejection of real systems and
  acceptance of false positives (Equation \ref{eq:nfp}).
  Diamonds incorporate initial maximum likelihood estimates of
  $d^{2}N/dzdW$ to adjust for variability in completeness across each
  bin (Equation \ref{eq:cdef_main}). Quoted errors reflect both
  counting statistics and uncertainty on completeness estimates, but
  the Poisson errors dominate in all cases.}
\label{fig:study_dndz_effects}
\end{figure}

Figure \ref{fig:study_dndz_effects} shows the effect of user errors on
the resultant $dN/dz$ for the rest frame equivalent width range
$W>0.3\textrm{\AA}$.  As previously stated, the triangles 
employ neither Equation \ref{eq:nfp} above to adjust for accepted false
positives, nor the full completeness $C_q(z,W)$,
which accounts for user rejection of real \mgii systems.  The squares
include both of these user error corrections.  The rejection of real
\mgii systems is by far the more dominant of these two effects, with
the correction for accepting false positives negligible compared to
the Poisson error bars (indicative of conservative users hesitant to
accept anything suspect).  The correction is the largest in the lowest redshift bin, where the
effective sensitivity is $\wminefflow \textrm{\AA}$ (as shown later), the poorest regime
in this survey.  As the
sensitivity improves with redshift, the user adjustments decrease in
magnitude.  Across the full range of redshifts and equivalent widths, the user accepts $\gtrsim 90\%$ of all real \mgii
systems.  The presented error bars include completeness errors, but
at $\lesssim 5\%$, these are subdominant to the Poisson errors.

\subsection{d$^{2}$N/dzdW}
\label{sub:d2ndzdw}

\begin{figure}
\epsscale{1.0}
\plotone{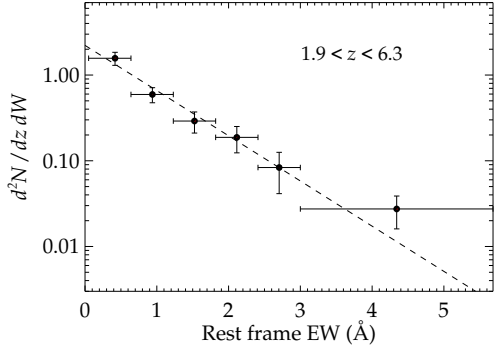}
\caption{The \mgii equivalent width distribution $d^2N/dzdW$ for the
  full survey, $1.9<z<6.3$. The overplotted line represents the
  maximum likelihood fit to the exponential form of Equation
  \ref{eq:mle_exp}, with best-fit parameters given in Table
  \ref{tab:dndws_var}.}
\label{fig:dndws}
\end{figure}

Figure \ref{fig:dndws} displays the rest frame equivalent width
distribution $d^2N/dzdW$ for the entire survey with all appropriate
corrections applied. Table \ref{tab:dndws_var} contains these calculated values. The distribution is fit well
by a simple exponential at $W\lesssim3.0\textrm{\AA}$, but shows a
relative overabundance of systems in the highest equivalent width bin.

\input{dndws_var}

\input{wstars}

Using maximum-likelihood techniques, we fit the equivalent width
distribution using the functional form
\begin{equation}
\frac{d^2N}{dzdW}=\frac{N_{*}}{W_{*}}e^{-W/W_{*}}.\label{eq:mle_exp}
\end{equation}
We first
calculated a maximum likelihood estimate of $W_{*}$, and then
calculated $N_{*}$ by insisting that the pathlength-weighted integral
of this population distribution equal the number of found systems,
\begin{equation}
\breve{N}_{k}\equiv\int\int g(z,W)\frac{d^2N}{dzdW}dzdW. \label{eq:nstarDef}
\end{equation}
This process yielded values of $W_{*}=\ewstar\pm\ealtvar\textrm{\AA}$
and $N_{*}=\enstar\pm\enstarerr$. Figure \ref{fig:dndws} displays this
maximum likelihood fit plotted over the binned, completeness-corrected
values.

We can use this result to calculate an effective equivalent width
sensitivity for the full survey. Suppose one assumes that the survey's
limiting equivalent width is characterized by a single number that is
constant at all redshifts and along all sightlines, $W_{\rm floor}$.
Then according to
Equation \ref{eq:gz_old},
\begin{equation}
g(z,W)\equiv\theta(W-W_{\rm floor})\sum_{q=1}^{Q}R_{q}(z).\label{eq:gz_weff}
\end{equation}
In this regime, the maximum likelihood estimate of $W_{*}$ may be
solved analytically as
\begin{eqnarray}
W_{*} & = & \left(\frac{1}{M}\sum_{m=1}^{M}W_{m}\right)-W_{\rm floor}\label{eq:wfloor}\\
 & \equiv & \overline{W}-W_{\rm floor},
\end{eqnarray}
where $\overline{W}$ is the average rest frame equivalent width of our
sample \citep{murdoch1986}. We define the effective sensitivity
$W_{\rm eff}$ of our survey as the value of $W_{\rm floor}$ in
this equation that, given our calculated value of $W_{*}$, would lead
to the average rest frame equivalent width $\overline{W}$ that we
observe.  For our reported values of $W_{*}$ and $\overline{W}$, we
obtain an effective sensitivity of
$W_{\rm eff}=\wmineff\pm\wminefferr\textrm{\AA}$ across the full
survey.

\begin{figure}
\epsscale{1.0}
\plotone{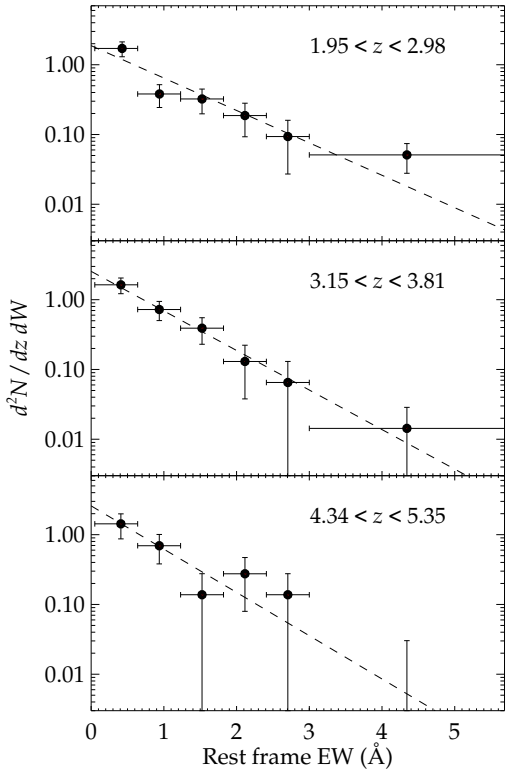}
\caption{The equivalent width distribution $d^{2}N/dzdW$ broken into
  three redshift intervals. Overplotted are the MLE fits to the
  exponential form of Equation \ref{eq:mle_exp}.  Table
  \ref{tab:dndws_var} contains the redshift ranges used and best-fit
  parameters for each case.  }
\label{fig:dndws_var}
\end{figure}

\begin{figure}
\epsscale{1.0}
\plotone{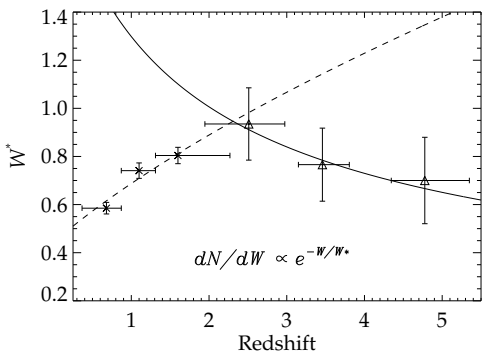}
\caption{Evolution in steepness of the equivalent width distribution,
  as parametrized by its exponential scale factor $W^*$.  The
  triangles represent maximum-likelihood estimates from the FIRE
  survey for the three redshift ranges in Figure \ref{fig:dndws_var}.
  The crosses represent the values from \citet{nestor2005}, determined
  from SDSS.  The overplotted solid line is the maximum likelihood
  fit for this survey's data from Equation \ref{eq:mle_wz}, and the
  dashed line is the analogous fit from \citet{nestor2005}.  $W^*$
  evolves significantly over redshift, and a simple polynomial form
  does not adequately capture this evolution.}
\label{fig:wstar_zs}
\end{figure}

In Figure \ref{fig:dndws_var}, we subdivide $d^{2}N/dzdW$ into three
different redshift ranges; $1.95<z<2.98$, $3.15<z<3.81$, and
$4.34<z<5.35$. Table \ref{tab:dndws_var} lists the values plotted.  We
also refit MLE estimates of $W_*$ for each of these regions
($\ewstarlow\pm\ealtvarlow$, $\ewstarmid\pm\ealtvarmid$, and
$\ewstarhigh\pm\ealtvarhigh$, respectively), and recalculated the effective rest
frame equivalent width sensitivities in each region
($\wminefflow\pm\wminefferrlow$, $\wmineffmid\pm\wminefferrmid$, and
$\wmineffhigh\pm\wminefferrhigh$, respectively).  In Figure
\ref{fig:wstar_zs} we plot $W_*(z)$, and include previously published
values from \citet{nestor2005}, to connect our results with lower
redshifts.  Table \ref{tab:wstars} contains a summary of all these values.

Clearly the trend of increasing $W^*$ seen at lower redshift does not
extrapolate to earlier epochs.  Rather, it exhibits a peak value of
$\sim 0.9$ at $z\sim 2.5$ and smaller values at both higher and lower
redshifts.  A larger value of $W^{*}$ corresponds to a flatter
equivalent width distribution with a larger fractional contribution
from high-$W_r$ absorbers.  It appears that going back in time,
$d^2N/dWdz$ evolves to include more strong \mgii systems until a
maximum is reached at $z\sim 2.5$, after which the relative frequency of
strong systems once again declines.  It is tempting to associate this
rise and fall of the strong \mgii systems with the concurrent rise and
fall of the global star formation rate; we will explore this
association in detail below.

To quantify $W_*$'s redshift evolution, we followed the lead of
\citet{nestor2005} and performed a MLE fit to
\begin{equation}
\frac{d^{2}N}{dzdW}=\frac{N_{*}}{W_{*}(1+z)^{\delta}}e^{-W/W_{*}(1+z)^{\delta}}.\label{eq:mle_wz}
\end{equation}
The best fit parameters for this from were
$W_{*}=\zewstar\pm\zealtvarone\textrm{\AA}$,
$\delta=\zedelta\pm\zealtvarfour$, and
$N_{*}=\zenstar\pm\zenstarerr$. The parameters $W_{*}$ and $\delta$
were anti-correlated, with $\rho_{W_{*}\delta}=\zealtvartwo$.  In Figure
\ref{fig:wstar_zs} we overplotted this best fit with a solid line, as well as the analogous MLE fit
of \citet{nestor2005} to his data with a dashed line.  The results show that a simple power law in
$(1+z)$ does not adequately capture $W_*$'s redshift evolution over
the full redshift range now available.  Rather, there is $2\sigma$
evidence that the overall distribution is steepening towards higher
$z$.

To examine evolution in the overall normalization of the frequency
distribution, we also fit our data to the functional form
\begin{equation}
\frac{d^{2}N}{dzdW}=\frac{N_{*}}{W_{*}}(1+z)^{\beta}e^{-W/W_{*}}.\label{eq:mle_pe}
\end{equation}
This exercise yielded $W_{*}=\pewstar\pm\pealtvarone\textrm{\AA}$,
$\beta=\pebeta\pm\pealtvarfour$, and $N_{*}=\penstar\pm\penstarerr$.
As is usual, these multi-dimensional MLE fits yield large errors for
small samples such as ours, but within these limitations we do not see
statistically significant evidence for redshift evolution in
the normalization of number counts. 

\subsection{dN/dz and dN/dX}
\label{sub:dndz}

\begin{figure}
\epsscale{1.0}
\plotone{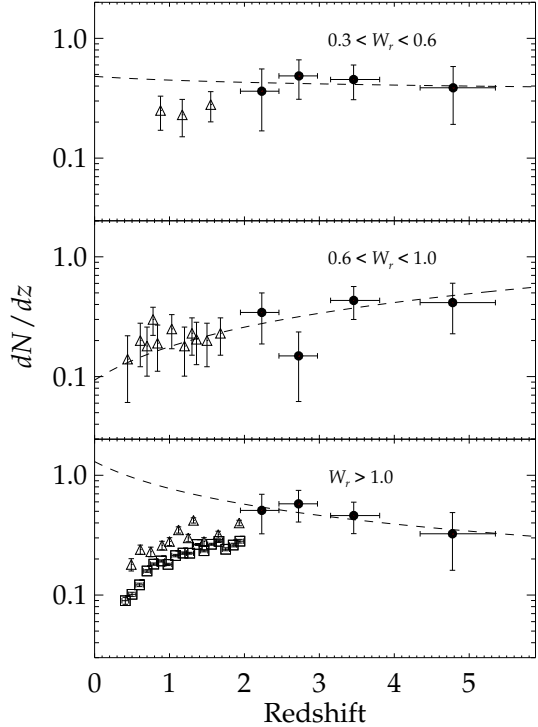}
\caption{The absorber line density $dN/dz$ for three different rest
  frame equivalent width ranges. The circular data points are from
  this study, the triangle data points are from \citet{nestor2005},
  and the square data points (bottom panel only) are from
  \citet{prochtersdss2006}. Table \ref{tab:dndzs} contains the values
  associated with the FIRE data.  The overplotted dashed lines
  represent maximum likelihood fits to the functional form given in
  Equation \ref{eq:MLE_dNdz}, over the range of the FIRE data only.}
\label{fig:dndzs}
\end{figure}

Figure \ref{fig:dndzs} displays the completeness-corrected, binned
values of $dN/dz$ for our \mgii sample in three different rest frame
equivalent width ranges. These ranges were chosen for easy comparison
with published SDSS \mgii surveys at lower redshift
\citep{nestor2005,prochtersdss2006}, which are overplotted. Table
\ref{tab:dndzs} contains the values of the data points in this figure.

We also fit the absorber line density to a simple polynomial,
\begin{equation}
\frac{dN}{dz}=N_{*}(1+z)^{\beta}.\label{eq:MLE_dNdz}
\end{equation}
We calculated a maximum likelihood fit of $\beta$, and then calculated
$N_{*}$ by insisting that our redshift path density integrated against
this absorber cross section equal the total number of systems located,
\begin{equation}
\breve{N}_{k}\equiv\int\int g(z,W)\frac{1}{\Delta W}\frac{dN}{dz}dzdW,
\end{equation}
where $\Delta W$ is the total rest equivalent width range probed.  We
performed this MLE fit over our full survey, and over the three rest
frame equivalent width ranges from Figure \ref{fig:dndzs}.  Table
\ref{tab:betas} contains these MLE fits, as well as the analogous fits
from \citet{prochtersdss2006}.  These fits are also overplotted with
dashed lines in Figure \ref{fig:dndzs}.

\begin{figure}
\epsscale{1.0}
\plotone{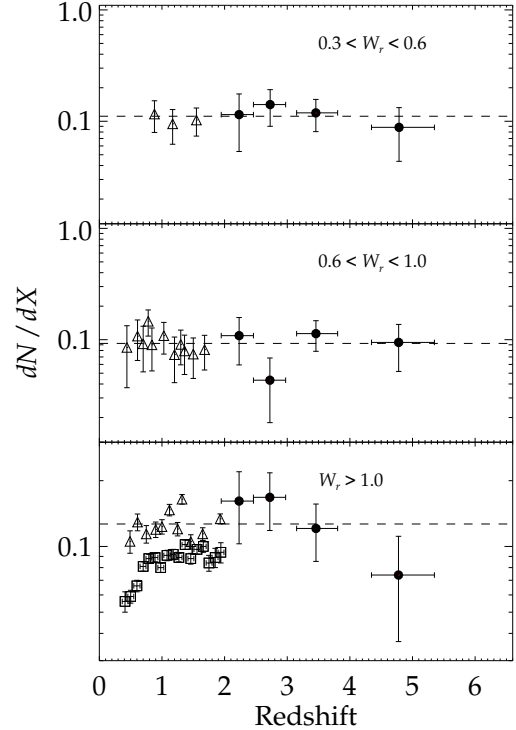}
\caption{Evolution of the line density per comoving length interval
  $dN/dX$. Overplotted horizontal lines represent the joint mean of
  the data from \citet{nestor2005} and this survey.  The top two
  panels, representing systems weaker than $W_r<1.0$\AA, are
  statistically consistent with zero evolution from $z=0.4$ to
  $z=5.5$.  In contrast, the stronger \mgii systems ($W_r >
  1\textrm{\AA}$) grow in number density until $z \sim 3$, after which
  the number density declines toward higher $z$.  }
\label{fig:dndxs}
\end{figure}

\input{dndzs}

\input{betas}

In addition, we calculated the comoving absorber line density
$dN/dX$. These completeness-corrected, binned results are shown in
Figure \ref{fig:dndxs}, with the corresponding results from
\citet{nestor2005} and \citet{prochtersdss2006} once again
overplotted.  The dashed horizontal lines show the straight mean
$dN/dX$ of our data combined with those of \citet{nestor2005}.  As in
the usual interpretation, a constant value of $dN/dX$ corresponds to a
population with no comoving evolution.  

The smaller rest frame equivalent width systems ($W_r<1\textrm{\AA}$)
show remarkably little evidence of cosmic evolution over the entire
range of redshifts probed by these surveys.  For the low end
$0.3<W_r<0.6$\AA ~sample the data are consistent ($\chi_\nu^2=0.2$)
with a constant value of $dN/dX=0.11$, and only $16\%$ standard
deviation.
The same is true
for $0.6<W_r<1.0$, where uncertainty from incompleteness is even
smaller.  In this $W_r$ range, $dN/dX=0.09$ with $\chi_\nu^2=0.5$ and
$25\%$ scatter around a constant value.  In contrast, the large rest
frame equivalent width systems ($W_r > 1\textrm{\AA}$) show
significant evolution over ours and previous surveys; $\chi_\nu^2=3.3$
for a constant value of $dN/dX$ over our survey and Nestor's, and
increases to $\chi_\nu^2=20$ if Prochter's $W_r>1.0$\AA ~data are
further included.  Table \ref{tab:dndzs} contains the values of our
new data points from this figure.

\section{Discussion}
\label{sec:discuss}

The initial results of our 46 sightline survey demonstrate that \mgii
absorbers were already commonplace at $z>5$.  Rather remarkably, the
comoving number density of \mgii absorbers with $W_r<1$ \AA ~is
completely flat at $dN/dX\sim 0.1$ with scatter of only $\sim 25\%$
between $z=0.4$ and $z=5$, a period of $>8$ Gyr.  In contrast, the
stronger \mgii absorbers grow in number density by a factor of $\sim
2-3$ from the present towards $z\sim 3$, after which they drop off by
a comparable amount toward $z\sim 5$.  The differential nature of the
evolution between strong and weak systems is reflected in the
equivalent width distribution, which is more shallow (higher $W^*$) at
$z\sim 3$ than it is at lower and higher redshifts.  The change in
{\em total} number counts is still small since the strong systems are
subdominant at all redshifts.

Local surveys of absorber-galaxy pairs typically find \mgii systems in
the extended $\sim 100$ kpc haloes of galaxies with $L\sim L^*$ or
slightly below \citep{bergeron_boisse,
  steidel,chen2010lowz,mgii_rotation,kacprzak2011incl}.  According to
the CDM picture of structure formation, galaxies resembling
present-day $L^*$ systems should be rare at $z\sim 4-5$.  Assuming
that \mgii absorbers are still found near galaxies at these epochs, it
follows that the typical galaxy giving rise to \mgii absorption in the
early universe must look quite different from the systems studied
locally.

The constancy of $dN/dX$ for weak absorbers implies that the product
$n\sigma$ of the comoving number density and physical cross section of
absorbing regions also remains constant.  A simple interpretation is
that these absorbers arise near the assembling progenitors of
present-day $L^*$ galaxies.  This would require that galaxy haloes
were populated with \mgii very early in their history, and that the
absorption properties were established before their stellar and ISM
components were fully formed.

A detailed analysis of absorption in individual \mgii systems will be
presented in a companion paper; however in a qualitative sense it
appears that \mgii systems at $z\sim 5$ are very similar to those at
low redshift, in contrast to the galaxy population.  This either
requires the absorption systems to persist in a steady state over much
of the Hubble time, or else be replenished periodically.  If \mgii
systems are replenished through stellar feedback, this process must
proceed in a way that does not produce a cumulative growth of the gas
radius and/or filling factor of \mgii.  This would seem to be a
challenge unless some portion of the gas re-accretes and cycles back
into the galaxy \citep{oppenheimer} or else the \mgii-bearing clouds
are out of equilibrium and transition to lower densities and higher
ionization states.

\subsection{\mgii and Cold Accretion Flows}

Recently \mgii has been suggested as a candidate tracer of ``cold''
accretion onto galaxies, since its relatively low ionization potential
(1.1 Ryd) does not allow for large \mgii fractions at high $T$.  Cold
accretion is thought to be the dominant mode of gas transport onto
galactic disks; it is particularly efficient in the early universe for
small, growing haloes \citep{keres, dekel, faucher_coldflows} and
could provide another ``renewable'' source for cool absorbers.  While
attractive, this model for the \mgii population faces some
complications at the high redshifts probed by our survey.
Principally, at these epochs, even though cold accretion is much
stronger \citep[$\dot{M}\propto (1+z)^{2.25}$;][]{dekel} and may
present a larger absorption cross section, the heavy element content
of the IGM gas reservoir is quite low: $Z/Z_\odot\sim 3\times 10^{-4}$
at $z\sim 4$ \citep{z4civ} and even lower at earlier times.  In
general, cold streams may manifest as metal-poor Lyman limit systems
\citep{fumagalli} more prominently than \mgii systems.  We may already
be seeing this phenomenon, in that the density of Lyman limit systems
and \mgii are of comparable magnitude at low redshift, but unlike
\mgii, the LLS density continues increasing with redshift
\citep{prochaska_lls, lls_survey} beyond $z>4$.

\citet{chenandtinker2009} present a model of the evolving \mgii
population which maps \mgii absorbers onto the dark matter halo mass
function, as a way of studying evolution of the underlying galaxies.
In their model, most \mgii absorbers arise in haloes with $M\sim
10^{11}-10^{12}M_\odot$ which are presumed to be growing via cold
accretion.  Shock heating reduces the supply of \mgii in larger mass
haloes, an effect which is required to match the observed
anti-correlation between galaxy mass and $W_r$ in the local universe
\citep{bouche2006anticorrelation,gauthier}.  However, an additional
suppression of absorption efficiency is also needed in low mass
($\lesssim 10^{11} M_\odot$) haloes to offset the rapidly rising halo
mass function in this regime.

\begin{figure}
\includegraphics[width=3.3in]{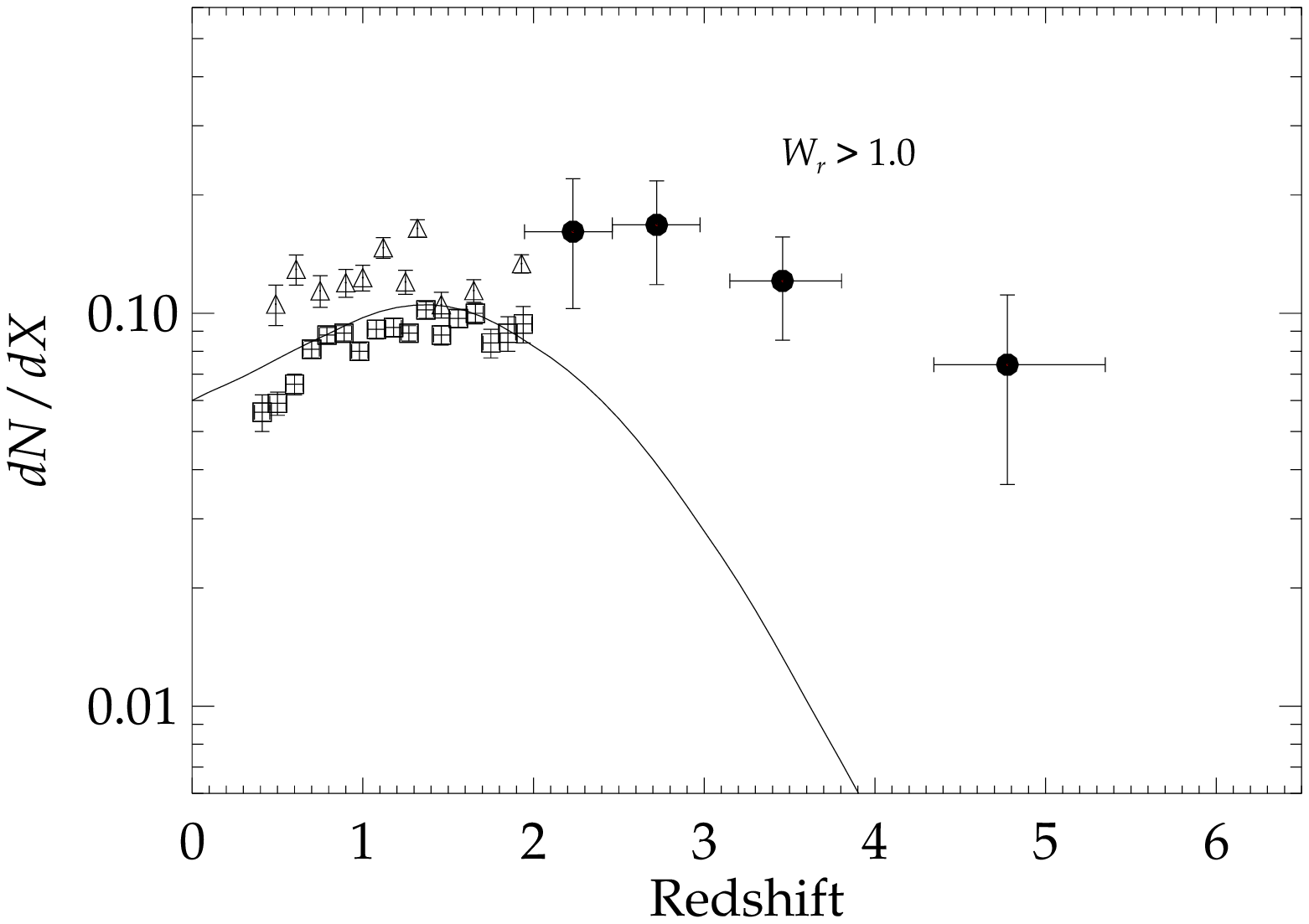}
\caption{The linear density $dN/dX$ for strong \mgii systems
  ($W_r>1.0\textrm{\AA}$) from \citet{prochtersdss2006} (squares),
  \citet{nestor2005} (triangles) and this survey (circles), compared
  with the linear density $dN/dX$ derived from the halo occupation
  distribution (HOD) model of \citet{chenandtinker2009} (solid line).
  The HOD model under-predicts the observed density at $z>3$ because
  of a paucity of large haloes at early times.  To narrow this
  discrepancy, the absorption efficiency of low-mass haloes would need
  to be enhanced in the early universe.}
\label{fig:tinker}
\end{figure}

This low-mass cutoff halts the overproduction of weak \mgii systems in
low mass haloes, aligning the model with observations at $z\sim 0.5$.
However at high redshift, a larger portion of the mass is concentrated
in these very haloes, most of which fall below the low-mass cut for
harboring \mgii absorption.  As a direct consequence, the simple
halo-occupation model under-predicts the measured $dN/dX$ at $z>2.5$
by orders of magnitude, as seen in Figure \ref{fig:tinker}.  It may be
possible to reconcile the model by boosting the absorption efficiency
of small mass haloes at early times, or else growing the absorption
cross-section per halo from $1/3$ of the virial radius at $z=0.5$ to
$>2 R_{vir}$ at early times.  However these solutions are not
physically motivated {\em a priori} and would need further
exploration.

It appears \mgii will be a difficult ion to use as a tracer of cold
accretion flows.  The competition between declining IGM metal
abundance and rising cold accretion cross section at earlier times
would require a fine balance to achieve the very flat trend we observe
in $dN/dX$.  While this solution is possible in principle, it will be
difficult to interpret.

\subsection{Comparison with Known Galaxy Populations}
\label{sec:dndxLum}

When comparing our \mgii number counts with galaxy populations,
another approach is to skip the halo mass function altogether, and
apply simple empirical scalings between \mgii gas halo size, covering
fraction, and galaxy luminosity to predict $dN/dX$.  The model cross
sections are integrated over measured luminosity functions as a
function of redshift.

To estimate each galaxy's cross section, we use the scalings of
\citet{chen2010lowz}, who studied a spectroscopic sample of 94
galaxy-absorber pairs at $\left<{z}\right>=0.2$.  Chen found that
galaxies with higher $B$ band luminosities possess more extended \mgii
absorbing haloes \cite[see also][]{steidel}, and fit the extent of
these gaseous haloes $R$ to a Holmberg-like scaling,
\begin{equation}
R(L_B) = R_0 \left(\frac{L_B}{L_{0}^*}\right)^{\beta}, \label{eq:rEqNoZ}
\end{equation}
where $R_0\approx 75 h^{-1}$ kpc, $\beta=0.35\pm0.03$, and $L_{0}^*$
is the luminosity associated with objects of $M_{B,0}^*=-19.8 +
5\log(h)$ at their survey redshift.  We use Equation \ref{eq:rEqNoZ}
together with measurements of the $B$ band luminosity function at
increasing redshift
\citep{lfHubble,lfCOMBO17,gabasch2004,lfDEEP2,marchesini2007} to
calculate the abundance and cross section of absorbers and compare
with our measurements of $dN/dX$.

When extrapolating Equation \ref{eq:rEqNoZ} to higher redshift, we
explored two ways of evolving the scaling.  In the simplest approach,
$R_0$ and $L^*_0$ are fixed at all redshifts, so galaxies with
$M_B=-19.8+5\log h$ {\em always} have haloes of radius $R_0$.
According to the studies cited above, $L_B^*$ evolves with redshift
toward brighter values in the past.  With a constant value of $L_0^*$
defining the halo scale, this implies that galaxies at $L_B^*$ will
have gas haloes increasingly larger than $R_0$ (i.e. $>75h^{-1}$ kpc)
in the more distant past.  For this reason, we will find it useful to
define the term $R^*(z)$, which denotes the gas radius of an $L_B^*$
galaxy at any redshift.

In the second approach, we associate gas haloes of size $R_0$ with
galaxies of $L=L_B^*(z)$ from the luminosity function at each epoch.
This is equivalent to setting $R^*(z)\equiv R_0$ and replacing $L_0^*$ with $L_B^*(z)$ in the notation above.  Since
$L_B^*$ increases with $z$, in this scenario galaxies at a fixed
luminosity would have smaller gas haloes in the past than in the
present day.

We next model the absorbers' covering fraction $\kappa$ as a function
of galaxy impact parameter $\rho$, using an empirical approximation
based on the results of \citet{chen2010lowz}:
\begin{equation}
\kappa(\rho,L_B,z)\equiv \left\{
\begin{array}{lll}
1 & \hspace{1em}\rho<a(L_B) \\
\frac{\rho - R(L_B)}{a(L_B) - R(L_B)} & \hspace{1em}a\leq\rho\leq R(L_B) \\
0 & \hspace{1em}\rho > R(L_B).
\end{array}\right.
\label{eq:kappa}
\end{equation}
Briefly, each galaxy has a central core of radius $a(L_B)$ within which
$\kappa=100\%$; between $a$ (the core radius) and $R$ (the halo
boundary) the covering fraction declines linearly to zero.  We assume
$a$ mimics $R$'s scaling with $L_B$ in both scenarios considered, but with $a_0=30 h^{-1}$ kpc.

The total absorption cross section $\sigma(L_B,z)$ presented by a
galaxy is then
\begin{eqnarray}
\sigma(L_B,z) & = & \int_0^{R(L_B,z)} 2\pi \rho \kappa(\rho,L_B,z) d\rho \\
& = & \left(\frac{L_B}{L^*_B(z)}\right)^{2\beta} \sigma^*(z). \label{eq:sigma}
\label{eq:cross}
\end{eqnarray}
Here $\sigma^*(z)$ is the cross section of an $L_B^*$ galaxy's gas
halo, which may evolve with redshift but does not depend on luminosity,
\begin{eqnarray}
\sigma^*(z) & \equiv & \bar{\kappa} \pi R^*(z)^{2} \\
\bar{\kappa} & \equiv & \frac{1}{3}\left( \frac{a_0^{2}}{R_0^{2}} + \frac{a_0}{R_0} + 1 \right) . \label{eq:sigmaStar}
\end{eqnarray}
The quantity $\bar{\kappa}$ is the average covering fraction of the
gaseous halo, and is $\bar{\kappa}=0.52$ with the given values of
$a_0$ and $R_0$.

\begin{figure}
\epsscale{1.15}
\plotone{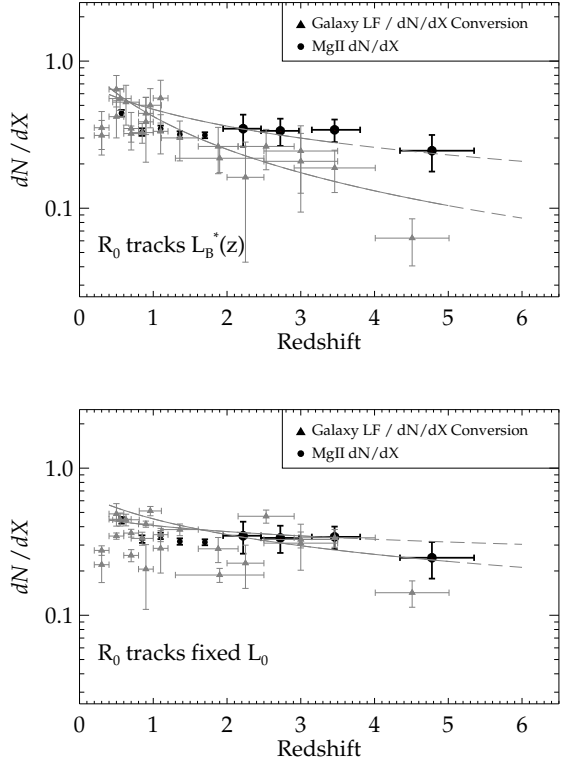} 
\caption{Linear densities $dN/dX$ from our survey (large, black
  circles) and from \citet{nestor2005} (smaller, black circles) for
  $W_r > 0.3\textrm{\AA}$, overplotted with $dN/dX$ derived with Equation \ref{eq:dndxLumAnalytic} using
  redshift-dependent $B$ band luminosity functions listed in the text.
  The top and bottom panels use different assumptions about the
  scaling of halo radius with luminosity: In the top panel, we assume the halo
  size $R_0$ is a fixed value and is associated with galaxies of
  $L_B^*$ at each epoch.  In the bottom panel, we assume $R_0$ is fixed, but
  associated with a fixed luminosity $L_0$ at all redshifts.  The
  smooth curves are derived from integrating redshift-dependent
  Schechter-function fits to the luminosity function; these are solid
  over the redshift ranges where measured, and high-redshift
  extrapolations are indicated with dashed lines.}  
\label{fig:schech}
\end{figure}

The \mgii frequency $dN/dX$ may be derived by combining this
per-galaxy cross section with the $B$ band luminosity function
$\phi(L_B,z)$:
\begin{equation}
\frac{dN}{dX} = \frac{c}{H_0}\int_{L_{\rm min}}^{\infty}
\sigma(L_B,z) ~\phi(L_B,z) ~dL_B , \label{eq:dndxLum}
\end{equation}
where $L_{\rm min}$ is a lower luminosity cutoff, which is not
specified {\em a priori}, but rather chosen to match the normalization
of the observed $dN/dX$.  Using a \citet{schechter1976} function
for $\phi(L_B,z)$,
we may calculate the linear density $dN/dX$ as
\begin{equation}
\frac{dN}{dX} = \frac{c}{H_0} \bar{\kappa} \pi R^*(z)^{2} \phi^*(z) \Gamma \left( \alpha(z) + 2 \beta + 1, \frac{L_{\hbox{min}}}{L^*_B(z)} \right), \label{eq:dndxLumAnalytic}
\end{equation}
where $\Gamma(s,x)$ is the upper incomplete gamma function.  This
expression is similar to one derived in \citet{churchill_weakmgii},
the only difference being the factor of $\bar{\kappa}$ correction for
partial covering.

We evaluated Equation \ref{eq:dndxLumAnalytic} using the $B$ band
luminosity functions of
\citet{lfHubble,lfCOMBO17,gabasch2004,lfDEEP2}; and
\citet{marchesini2007} to calculate $dN/dX$ from $z=0$ to $z=5$.  The
results are shown in Figure \ref{fig:schech}, with triangles and
circles representing the predicted and measured $dN/dX$ (for
$W_r>0.3$\AA), respectively.  The solid curves indicate the predicted
evolution based on redshift-dependent fits to the luminosity function
parameters from \citet{lfHubbleZdep} and \citet{gabasch2004}.

The top panel shows results for the model where the luminosity
associated with $R_0$-sized haloes tracks $L_B^*(z)$.  In this case,
the observed and predicted $dN/dX$ match best when $L_{\rm
  min}/L_B^*=0.024\pm0.005$.  However the model predicts a downward
evolution in $dN/dX$ with $z$ that is slightly steeper than what is
observed.  This means that the typical cross section per galaxy is
higher than the model predicts at high redshift, suggesting that either the haloes
are growing slightly in radius or their filling factors
are increasing.

The bottom panel shows the simple model where the fiducial luminosity
associated with $R_0$ haloes is fixed at all redshifts.  Our best fit
is obtained with $L_{\rm min}/L_B^*=0.2656\pm0.0139$, almost an order of
magnitude larger than the previous model.  The redshift dependence for this model is flatter,
in better agreement with our survey data.  The relatively high value
of $L_{\rm min}$ indicates that a large population of very low
luminosity galaxies need not be invoked to explain the frequency of
\mgii systems in this model; for the most part, all $W_r>0.3$ \mgii
systems surround galaxies we can readily observe, even at high
redshift.  

This picture could be invalidated if significant numbers of \mgii
systems are discovered near galaxies with $L<0.2L_B^*$.  However to
date such dwarf-\mgii associations have not been established except in
a few cases.  Even in \citet{chen2010lowz}'s sample of 94 pairs, $<10\%$ of the
galaxies meet this criterion.  They are predominantly at very small
impact parameter ($\sim 10h^{-1}$ kpc) and even then show only $\sim
50\%$ coverage (in contrast to our model, in which the coverage is $100\%$ out to $\sim 30 h^{-1}$ kpc).

The $dN/dX$ points in Figure \ref{fig:schech} apply for $W_r>0.3$\AA
~since this was the range over which the $R(L)$ correlation was first
determined.  However we have also seen that if we restrict further to
$W_r>1.0$\AA, $dN/dX$ evolves to a peak at $z\sim 3$ and declines
thereafter.  Neither of our halo models is able to reproduce this
feature, which indicates that other physics must be at play.  If the
systems giving rise to the strong absorbers are contained in the
$B$ band luminosity function, then their gas cross section and/or
covering fraction must rise and then fall again.  Alternatively they
may reflect systems whose comoving number density rises and falls
without being picked up in rest frame $B$ band surveys.

Rather than start with a halo and coverage model and derive $dN/dX$,
we may instead start with the observed \mgii $dN/dX$ and estimate the
effective size of gaseous haloes.  To do this, we define the effective
halo radius $R^{\rm eff}(z)$ of an absorber according to
\begin{eqnarray}
\pi R^{\rm eff}(z)^2 & \equiv & \pi \kappa_L(z) R_L(z)^2 \\ & = &
\frac{H_0}{c} \frac{1}{n(z,L_{\hbox{min}})} \frac{dN}{dX}
\label{eq:reffNew}
\end{eqnarray}
where $\kappa_L(z)$ is the luminosity-weighted covering fraction and
$R_L(z)$ is luminosity-weighted gaseous halo radius.

\begin{figure} 
\includegraphics[width=3.3in]{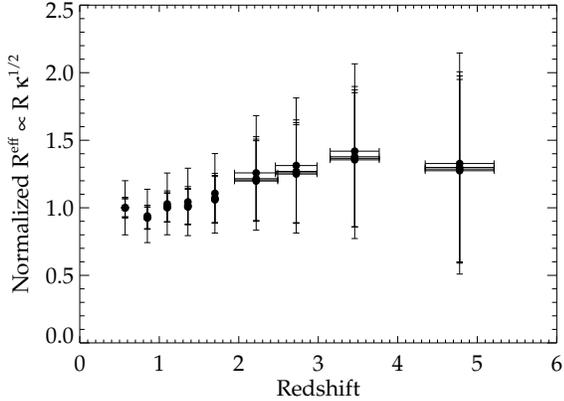} 
\caption{Effective
halo radius (Equation \ref{eq:reffNew}) as a function of redshift for
the observed \mgii $dN/dX$ of this study and from \citet{nestor2005}, normalized by its value in the lowest redshift bin.
We derived the absorber number density $n(z,L_{\hbox{min}})$ from a
combination of the redshift dependent luminosity function models of
\citet{lfHubbleZdep} and \citet{gabasch2004} for four lower luminosity
cutoffs: 0.001, 0.01, 0.1 and 1.0 $L_B^*(z)$.  With all four choices, the
effective halo radius increases with redshift by about $\sim 40\%$ from today until $z\sim 3-4$, and
then decreases again (although this decrease is not statistically significant given our error bars). } \label{fig:reff} 
\end{figure}

In Figure \ref{fig:reff}, we plot the effective halo radius as a
function of redshift, normalized by the radius calculated in the lowest redshift bin.  We use an average of the redshift evolution
luminosity function models of \citet{lfHubbleZdep} and
\citet{gabasch2004} in conjunction with four different lower
luminosity cutoffs: 0.001, 0.01, 0.1, and 1.0 $L_B^*(z)$.  All four
choices of a lower luminosity cutoff result in the effective absorbing
halo radius $R^{\hbox{eff}}(z)$ increasing by $\sim 40\%$ from today to
$z\sim 3-4$, and then slightly decreasing again (although the error bars are too large to consider this final decrease statistically significant).  This increase in effective halo
radius must result from either an increase in the luminosity-weighted
covering fraction $\kappa_L(z)$ at earlier times, or an increase in
the extent of the gaseous absorbing haloes $R_L(z)$ at earlier times.  Although the lower
luminosity cutoff still remains a free parameter in this model, the strong
quantitative agreement of the normalized effective halo radii across three orders of magnitude in
$L_{\hbox{min}}$ suggests this feature is largely independent of the
lower luminosity cutoff employed.

\subsection{Deriving the SFR Density from dN/dX}
\label{sec:SFR}

\begin{figure}
\includegraphics[width=3.3in]{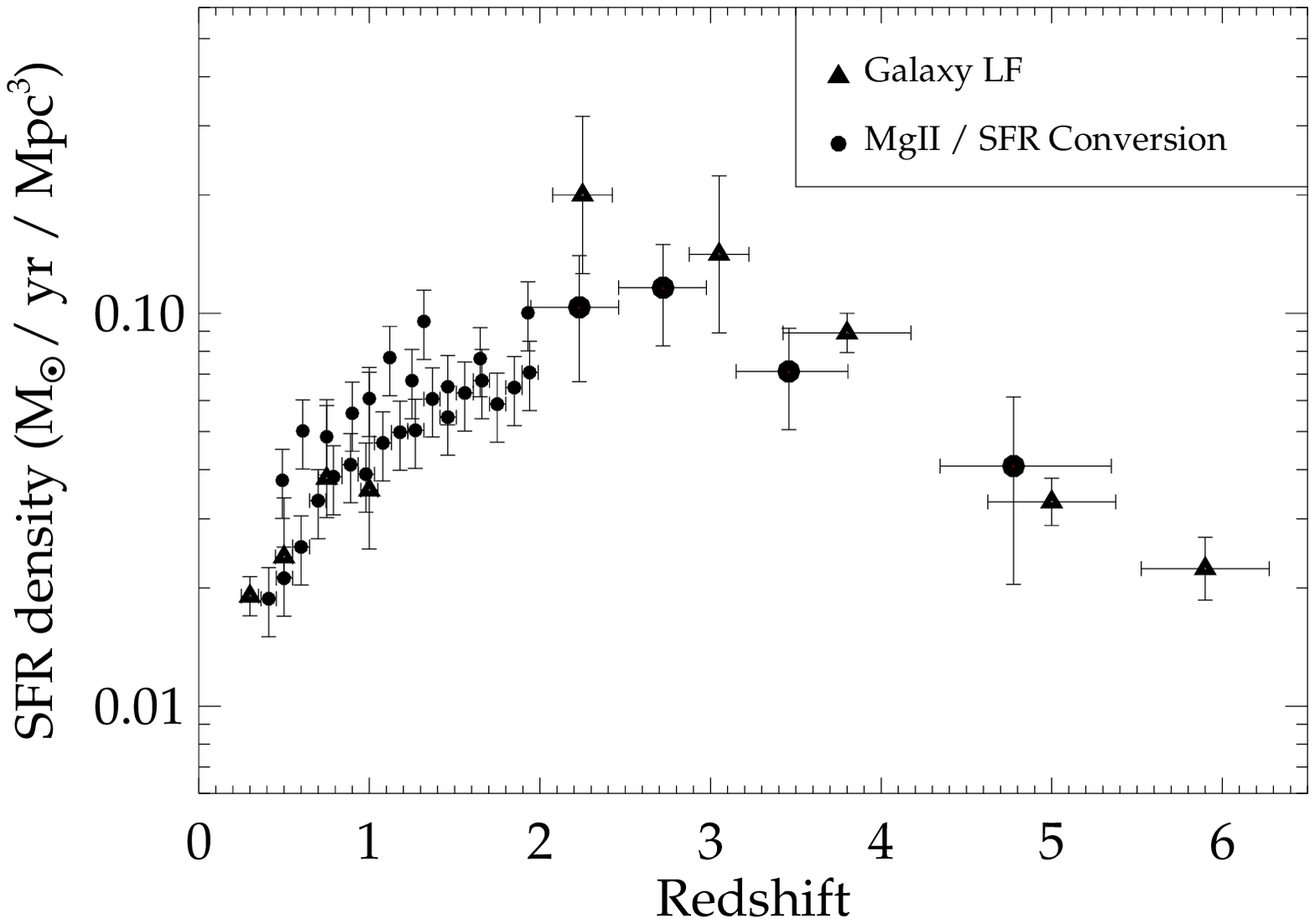}
\caption{The star formation rate density $\dot{\rho}_{\star}$ derived
  using the method of \citet{menardo2} on large \mgii systems
  ($W_r>1\textrm{\AA}$) from this survey, \citet{nestor2005}, and
  \citet{prochtersdss2006} (circles; larger circles are this survey).
  Recent star formation rate density calculations derived from
  observations with the Hubble Telescope in \citet{bouwens2010} and
  \citet{bouwens2011} are overplotted with triangles.  The method of
  \citet{menardo2} produces good qualitative agreement with the Hubble
  results, although the error bars are quite large.  Applying this
  \mgii conversion to weaker \mgii systems ($W_r<1\textrm{\AA}$) does
  not lead to a match with these Hubble rates.}
\label{fig:sfr}
\end{figure}
Thus far we have focused the discussion on moderate strength \mgii
absorbers which do not evolve with redshift.  But the systems with
$W_r>1$\AA ~{\em do} evolve, with a characteristic rise and fall that
suggests a possible connection to the cosmic star formation rate
density

This connection is also explored in \citet{menardo2}, where a method
is outlined for deriving the SFR density from the \mgii population
distribution by establishing a connection between \mgii rest frame
equivalent width $W_r$ and \oii luminosity, which previous studies
have shown may be used as a gauge of the SFR
\citep{gallagher1989,kennicutt1992,hopkinsOii,kewley2004,mouhcine2005}.
Since the detection of \mgii using spectroscopic absorption does not
depend on luminosity, distance, or dust extinction, such a probe would
represent a valuable asset in pinning down the star formation history
of the universe.

In their paper, the authors search for \oii emission lines from \mgii
host galaxies imprinted on a sample of QSO spectra containing over
8500 known \mgii absorbers at redshifts $0.36<z<1.3$.  They detect a
$15\sigma$ correlation between the absorber strength $W_r$ and the
median \oii luminosity surface density $\left<{\Sigma_{L_{OII}}}\right>$, which
they fit to the functional form
\begin{equation}
\left<{\Sigma_{L_{OII}}}\right> = A \left(\frac{W_r}{1\textrm{\AA}}\right)^\alpha, \label{eq:menardCorr}
\end{equation}
where $A=(1.48\pm0.18)\times 10^{37}$ ergs s$^{-1}$ kpc$^2$ and
$\alpha=1.75\pm0.11$.  The authors calculate the co-moving \oii
luminosity density $\mathcal{L}_{OII}(z)$ probed by \mgii absorbers
from this by using the completeness corrected population distribution
from \citet{nestor2005} according to
\begin{equation}
\mathcal{L}_{OII}(z) = \int dW \frac{d^2N}{dWdX} \Sigma_{L_{OII}}.
\end{equation}
Finally, they convert this luminosity density $\mathcal{L}_{OII}(z)$
to a luminosity $L_{OII}$ and derive a SFR density
$\dot{\rho}_{\star}$ by using an average over scalings from
\citet{zhuSFR},
\begin{equation}
\dot{\rho}_{\star} = 4.2\hbox{e-}41 \times L_{OII} \hbox{ M}_\odot \hbox{ yr$^{-1}$ Mpc$^{-3}$}.\label{eq:ks}
\end{equation}
Using this technique, \citet{menardo2} convert the data from
\citet{quiderSDSS} to SFR densities, and then overplot SFR density
data provided in \citet{hopkinsSFR}.  The results show agreement up to
$z=2.2$.

More recently, \citet{lopezChen2011} have argued that the
$W_r$-$L_{[OII]}$ correlation reported in \citet{menardo2} does not
reflect a causal connection between \mgii absorption and star
formation.  Rather, they demonstrate that the observed correlation
could plausibly arise from a combination of the empirically observed
decline in $W_r$ at increasing galactocentric radius, and
differential loss of galaxy light from the finite fiber aperture of
the SDSS spectrograph.  The authors run a large Monte Carlo simulation
which reproduces the $W_r$-$\left<{\Sigma_{L_{OII}}}\right>$
relationship found in \citet{menardo2}.

Nevertheless, in light of the observed evolution in $dN/dX$ for $W_r >
1$\AA, the large, likely non-gravitational velocity spreads observed
for strong \mgii systems (Figure \ref{fig:vspreads}), and the
demonstrated connection between the strongest \mgii absorbers and star
forming galaxies \citep{bouche2007ha,noterdaeme2010,nestor2010strong},
it is tempting to explore Menard's method, while keeping in mind its
possible limitations.  The broad redshift range of our sample,
combined with those of the SDSS studies, puts us in unique position to
compare with the cosmic star formation history.

In Figure \ref{fig:sfr}, we show the SFR density calculated using
Equations \ref{eq:menardCorr} to \ref{eq:ks}, for \mgii absorbers
having $W_r > 1$ \AA ~(large circles).  We also include previous results at low
redshift from \citet{nestor2005} and \citet{prochtersdss2006} over the
same equivalent width range (smaller circles).  For comparison, we overplot (with triangles) star
formation rate densities obtained using recent determinations based on
dropout counts from deep HST/WFC3 galaxy surveys at $z > 4$
\citep{bouwens2010,bouwens2011}, from a combination of ground-based spectroscopic $H\alpha$ and Spitzer MIPS 24$\mu m$ data at $z\sim
3$ \citep{reddysteidel}, and from the GALEX VIMOS-VLT Deep Survey at lower redshifts \citep{arnouts}.  Menard's
transformation produces star formation rate densities in excellent
agreement with the galaxy survey literature throughout the entire
\mgii absorption redshift search range probed, $0.5\lesssim z
\lesssim5$.  These results do not conclusively demonstrate the
accuracy of Equation \ref{eq:menardCorr}---indeed application of
Menard's redshift dependent version of Equation \ref{eq:menardCorr}
led to very high SFR values per absorber at $z\sim5$ and therefore
predicted significantly higher SFR densities.  However this exercise
certainly suggests that the connection between \mgii absorption and
SFR is real for the strongest subset of \mgii absorbers, and stretches
over the full cosmic history.

Application of the \mgii-SFR conversion to weaker absorbers
($W_r<1\textrm{\AA}$) does not produce good agreement with galaxy
surveys.  These systems, as seen in Figure \ref{fig:dndxs}, exhibit
remarkably little evolution in $dN/dX$ over all redshifts considered
($0.5\lesssim z \lesssim5$).  This suggests that any physical
connection between \mgii and star formation is mostly limited to the
stronger end of the \mgii equivalent width distribution.  Indeed
different astrophysical processes may govern the evolution of systems
at high and low equivalent width, with the division occurring roughly
at $W_R\sim 1$\AA ~as suggested by \citet{lovegrove2011} and
\citet{kacprzak2011schech}.

\section{Conclusions}
\label{sec:conclusion}

Using FIRE, we have conducted an infrared survey of intervening \mgii
absorption toward the sightlines of 46 distant quasars with redshifts
$3.55\leq z \leq6.28$.  The overall survey has an effective
sensitivity of $\wmineff\pm\wminefferr\textrm{\AA}$ and a FWHM
resolution of $50$ km/s.  A combination of automated and human methods
were used to identify and measure the properties of detected systems,
and we performed extensive simulations to characterize our sample's
incompleteness and contamination from false positives.  Our main
findings are as follows:
\begin{enumerate}
\item We find 110 isolated \mgii systems (and one proximate system)
  ranging in rest equivalent width from $W_r=0.08\textrm{\AA}$ to
  $W_r=5.58\textrm{\AA}$ and in redshift from $z=1.98$ to $z=5.33$,
  including 5 systems with $z>5$.
\item The population distribution $d^2N/dzdW$ resembles an exponential
  with a characteristic scale $W_{*}=\ewstar\pm\ealtvar\textrm{\AA}$
  across the full survey.  When we divide the survey into three
  redshift regions and combine our results with earlier results from
  \citet{nestor2005}, we find that this characteristic scale $W_{*}$
  rises with redshift until $z\sim 2-3$ (where we find more strong
  systems), and then falls towards higher redshifts.
\item For weaker \mgii absorbers ($W_r<1$ \AA), the linear density is
  statistically consistent with no evolution, having $dN/dX\sim 0.1$
  with a scatter of only $\sim 15\% - 25\%$ from $z=0.4$ to $z>5$, a period of
  $>8$ Gyr.  In contrast, the stronger \mgii absorbers ($W_r>1$ \AA)
  grow in number density by a factor of $\sim 2-3$ from the present
  towards $z\sim 3$, after which they drop off by a comparable amount
  toward $z\sim 5$.  We interpret this as evidence of two distinct
  \mgii populations, one which follows the global star formation rate
  density, and another which is nearly constant in comoving aggregate
  cross section.
\item For strong \mgii absorbers ($W_r>1$ \AA), the \mgiinsp-to-star
  formation rate scaling of \citet{menardo2}, together with our
  measured \mgii frequency, agrees remarkably well with classically
  measured star formation density rates.  While the recent analysis of
  \citet{lopezChen2011} cautions against a detailed exploration of
  this specific model, the excellent agreement with our $dN/dX$
  measurements are highly suggestive of a connection between strong
  \mgii absorption and star formation.
\item For the weaker systems, models based on the predicted \mgii
  occupation of dark matter haloes substantially under-predict $dN/dX$
  at high redshifts.  This discrepancy derives from the low absorption
  efficiency of small mass haloes locally (which are used for
  scaling), combined with the relative scarcity of high mass haloes in
  the early universe.  To resolve the difference, the numerous
  low-mass haloes present at early times would need to have an
  enhanced absorption efficiency relative to present day levels.
\item We are able to produce the observed $dN/dX$ more accurately
  using scaling relations between galaxy luminosity and halo size
  derived at low redshift, but integrated over appropriate high
  redshift determinations of the galaxy luminosity function.  The
  correspondence is most accurate when the \mgii absorbing halo's size
  scales with a fixed, redshift-independent fiducial luminosity.
  Since $L_B^*$ increases with redshift, this implies that the
  effective gaseous halo of an $L_B^*$ galaxy is larger at earlier
  times.  This model leaves the lower integration limit $L_{min}$ down
  the faint-end slope as a free parameter, set to match the
  normalization of $dN/dX$.  Our best match is for $L_{min}\approx
  0.2L_B^*$, so that large populations of low-luminosity galaxies are
  not strictly required to explain the observed \mgii frequency, even
  at high redshift.
\end{enumerate}

Our results indicate that \mgii absorption remains commonplace in the
early universe, having evolved relatively little over a period
exceeding $8$ Gyr.  This is in contrast to the \civ number counts,
which decline by almost an order of magnitude over the same interval
(K. Cooksey, private communication), and the density of the
intergalactic \hi Lyman alpha forest, which thickens markedly toward
the epoch of reionization \citep{becker_hi}.

At low redshift, \mgii absorbers are typically associated with $\sim
100$ kpc gas haloes surrounding $\sim L^*$ galaxies.  The \mgii counts
at high redshift can also plausibly be reproduced using fairly
luminous galaxy populations.  As will be shown in forthcoming work,
the \hi and other metal line properties of the high-redshift \mgii
systems also do not differ dramatically from the local population.

Yet the properties of the host galaxies themselves must have evolved
substantially over the same period.  As estimated by
\citet{gabasch2004} $L_B^*$ brightens by a factor of $\sim 2$, while
$\Phi_*$ declines by an order of magnitude over the redshift path
where \mgii has now been characterized.  If the high redshift \mgii
absorbers are similarly associated with distinct galaxies, then the
mass substructure within their dark matter haloes must have evolved
significantly from $z\sim 5$ to the present.  Moreover, high
resolution, rest-frame optical morphology studies of distant galaxies
generally find that at fixed stellar mass, galaxies have smaller half
light radii ($0.7-3$ kpc at $z\sim 3$) going backwards in time
\citep{law_wfc3,papovich,franx,vandokkum}.

It appears that as the stellar populations of typical galaxies grow
from the inside out, the gaseous haloes of the same galaxies may be
populated with metals very early on, and that the gas halo properties
evolve much more weakly than do the galaxies themselves.  If our
hypothesis is correct that the observed $dN/dz$ may be recovered
entirely by galaxies with $L>0.2L^*$, then it should be possible to
observe such systems directly at intermediate redshifts, up to $z\sim
3-4$.  Along with further spectroscopy to improve upon the modest
sample sizes presented here, such follow up of \mgii-selected galaxies
may constitute a fruitful means for studying the concurrent buildup of
the stellar and near field circum-galactic environments.

\acknowledgments

It is a pleasure to thank the staff of the Magellan Telescopes and Las
Campanas Observatory for their assistance in obtaining the data
presented herein.  We also acknowledge helpful conversations with
K. Cooksey, H.W. Chen, B. Menard during preparation of the
manuscript.  RAS acknowledges the gracious hospitality of J. Hennawi
and the MPIA/Heidelberg, where a portion of this work was completed.
RAS also recognizes the culturally significant role of the
A.J. Burgasser Chair in Astrophysics.  Finally, we gratefully
acknowledge financial support from the NSF under grant AST-0908920.

\section*{Appendix}
\subsection*{Mathematical Framework for Variable Completeness}
\label{sub:a1}

\renewcommand{\theequation}{A\arabic{equation}}
\setcounter{equation}{0}

If our data set were 100\% complete over the \mgii redshift search range
and rest frame equivalent width range $W_{k}^{\hbox{min}}<W<W_{k}^{\hbox{max}}$
spanned by a bin $k$, then the expected number of \mgii systems $N_{k}$
found within this bin would be given by the integral of the population
distribution $d^{2}N/dzdW$ weighted by the sum of the regions $R_{q}(z)$
probed along each sightline; more specifically,
\begin{equation}
N_{k}=\int\int_{k}\left(\sum_{q=1}^{Q}R_{q}(z)\right)\frac{d^{2}N}{dzdW}dzdW,\label{eq:N_100pct}
\end{equation}
where the range of the integral, denoted by subscript $k$, is over
the redshift and rest frame equivalent width range of the bin.
The total path $\Delta z_{k}$ in this bin would then be 
\begin{equation}
\Delta z_{k}=\int_{k}\left(\sum_{q=1}^{Q}R_{q}(z)\right)dz,\label{eq:path_100}
\end{equation}
where the integral is now only over redshift. If we denote the change
in equivalent width of the bin as $\Delta W_{k}\equiv W_{k}^{\hbox{max}}-W_{k}^{\hbox{min}}$,
then the population density is estimated as 
\begin{equation}
\left(\frac{d^{2}N}{dzdW}\right)_{k}=\frac{N_{k}}{\Delta z_{k}\Delta W_{k}}.\label{eq:pop_dens}
\end{equation}

With less than 100\% completeness, the expected number of \mgii systems
$\breve{N}_{k}$ found within this bin is instead
\begin{eqnarray}
\breve{N}_{k} & = & \int\int_{k}\left(\sum_{q=1}^{Q}R_{q}(z)C_{q}(z,W)\right)\frac{d^{2}N}{dzdW}dzdW\label{eq:Nbreve_comp} \\
 & = & \int\int_{k}g(z,W)\frac{d^{2}N}{dzdW}dzdW.\label{eq:N_lt100pct}
\end{eqnarray}
We define the proportionality constant between $N_{k}$ and $\breve{N}_{k}$
to be the average completeness $\overline{C}_{k}$ in this bin,
\begin{equation}
\breve{N}_{k}\equiv\overline{C}_{k}N_{k},\label{eq:cdef1}
\end{equation}
which leads to
\begin{equation}
\overline{C}_{k}\equiv\frac{\int\int_{k}\left(\sum_{q=1}^{Q}R_{q}(z)C_{q}(z,W)\right)\frac{d^{2}N}{dzdW}dzdW}{\int\int_{k}\left(\sum_{q=1}^{Q}R_{q}(z)\right)\frac{d^{2}N}{dzdW}dzdW}.\label{eq:cdef2}
\end{equation}
Mathematically, this is equivalent to calculating the average completeness
across the bin using a probability density proportional to the product
of the population distribution and the summed regions probed.

According to Equations \ref{eq:pop_dens} and \ref{eq:cdef1}
above, the completeness-corrected estimate of the population density
is then 
\begin{equation}
\left(\frac{d^{2}N}{dzdW}\right)_{k}=\frac{\breve{N}_{k}}{\overline{C}_{k}\Delta z_{k}\Delta W_{k}}.\label{eq:pop_dens_lt100}
\end{equation}
Similarly, the estimates of the rest frame equivalent width distribution
$dN/dW$ and the absorber line density $dN/dz$ are given by
\begin{equation}
\left(\frac{dN}{dW}\right)_{k}=\frac{\breve{N}_{k}}{\overline{C}_{k}\Delta W_{k}}\label{eq:dNdW_lt100}
\end{equation}
and
\begin{equation}
\left(\frac{dN}{dZ}\right)_{k}=\frac{\breve{N}_{k}}{\overline{C}_{k}\Delta z_{k}}.\label{eq:dNdz_lt100}
\end{equation}

If, as we argue in the main text, we assume that the population distribution
$d^{2}N/dzdW$ is constant over bin $k$ when calculating $\overline{C}_{k}$,
then the average completeness across the bin reduces to 
\begin{eqnarray}
\overline{C}_{k} & \approx & \frac{\int\int_{k}\left(\sum_{q=1}^{Q}R_{q}(z)C_{q}(z,W)\right)dzdW}{\int\int_{k}\left(\sum_{q=1}^{Q}R_{q}(z)\right)dzdW}\label{eq:c_k}\\
 & = & \frac{\int\int_{k}g(z,W)dzdW}{\Delta z_{k}\Delta W_{k}},
\end{eqnarray}
where we have employed the definition of $g(z,W)$ from Equation \ref{eq:gz_new}
and the definition of the total path length 
given in Equation \ref{eq:path_100}. Plugging this equation for
$\overline{C}_{k}$ back into Equation \ref{eq:pop_dens_lt100}
for the population density above yields
\begin{equation}
\left(\frac{d^{2}N}{dzdW}\right)_{k}=\frac{\breve{N}_{k}}{\int_{k}g(z,W)dzdW}.\label{eq:pop_dens_approx-1}
\end{equation}

\subsection*{Mathematical Framework for False Positive Correction}
\label{sub:a2}

Let $d^{2}F/dzdW$ denote the population density of false positives
that our automated line identification system flags as potential candidates.
The total number of false positives $F_{k}$ in bin $k$ is given
by
\begin{equation}
F_{k}=\int\int_{k}\left(\sum_{q=1}^{Q}R_{q}(z)\right)\frac{d^{2}F}{dzdW}dzdW.\label{eq:Fk}
\end{equation}
If $A_{q}^{\hbox{F}}(z,W)$ is
the probability of the user accepting a false positive in sightline
$q$ at redshift $z$ and rest frame equivalent width $W$, then the
number of actual false positives $\breve{F}_{k}^{C}$ that the user
correctly identifies is 
\begin{equation}
\breve{F}_{k}^{C}=\int\int_{k}\left(\sum_{q=1}^{Q}R_{q}(z)\left(1-A_{q}^{\hbox{F}}(z,W)\right)\right)\frac{d^{2}F}{dzdW}dzdW.\label{eq:FkC}
\end{equation}
Along with these, the user incorrectly identifies some real \mgii systems
as false positives,
\begin{equation}
\breve{F}_{k}^{I}=\int\int_{k}\left(\sum_{q=1}^{Q}R_{q}(z)\left(L_{q}(z,W)-C_{q}(z,W)\right)\right)\frac{d^{2}N}{dzdW}dzdW,\label{eq:FkI}
\end{equation}
where $L_{q}(z,W)$ is the automated completeness and $C_{q}(z,W)$
is the total completeness as given in Equation \ref{eq:comp_tot_q},
which also accounts for the user's decisions.

In terms of accepted candidates, the number of correctly accepted
\mgii candidates $\breve{N}_{k}^{C}$ follows from Equation \ref{eq:Nbreve_comp},
\begin{equation}
\breve{N}_{k}^{C}=\int\int_{k}\left(\sum_{q=1}^{Q}R_{q}(z)C_{q}(z,W)\right)\frac{d^{2}N}{dzdW}dzdW,\label{eq:NkC}
\end{equation}
Along with these systems,
the user incorrectly accepts some false positives, the number $\breve{N}_{k}^{I}$
of which is given by
\begin{equation}
\breve{N}_{k}^{I}=\int\int_{k}\left(\sum_{q=1}^{Q}R_{q}(z)A_{q}^{\hbox{F}}(z,W)\right)\frac{d^{2}F}{dzdW}dzdW.\label{eq:NkI}
\end{equation}

As before, we will assume that the \mgii and false positive population
densities vary negligibly over the course of a single bin. Therefore,
we may define the expected value of an arbitrary function $h(z,W)$
over bin $k$ according to 
\begin{equation}
\overline{h}_{k}\equiv\frac{\int\int_{k}\left(\sum_{q=1}^{Q}R_{q}(z)h_{q}(z,W)\right)dzdW}{\int\int_{k}\left(\sum_{q=1}^{Q}R_{q}(z)\right)dzdW}.\label{eq:avgDef}
\end{equation}
With this definition, the number of identified (real and spurious)
\mgii systems $\breve{N}_{k}$ and false positives $\breve{F}_{k}$
in bin $k$ are given by 
\begin{eqnarray*}
\breve{N}_{k} & = & \breve{N}_{k}^{C} + \breve{N}_{k}^{I} = N_{k}\overline{C}_{k}+F_{k}\overline{A}_{k}^{\hbox{F}} \label{eq:NkU}  \\
\breve{F}{}_{k} & = & \breve{F}_{k}^{C} + \breve{F}_{k}^{I} = (1-\overline{A}_{k}^{\hbox{F}})F_{k}+N_{k}(\overline{L}_{k}-\overline{C}_{k}). \label{eq:FkU}
\end{eqnarray*}
Here, we have two equations and two unknowns, namely the number of
real systems $N_{k}$ and true false positives $F_{k}$. Solving for
$N_{k}$ gives
\begin{equation}
N_{k}=\frac{\breve{N}_{k}(1-\overline{A}_{k}^{\hbox{F}})-\overline{A}_{k}^{\hbox{F}}\breve{F}_{k}}{\overline{C}_{k}-\overline{L}_{k}\overline{A}_{k}^{\hbox{F}}}.\label{eq:N_user}
\end{equation}
The population densities are then given by Equations \ref{eq:pops_d2ndzdw}
to \ref{eq:pops_dndw} with this value substituted for $N_{k}$.

\bibliography{mgii}{}
\bibliographystyle{apj}

\clearpage

\end{document}

%% file: macros.tex
\newcommand{\civ}{C~{\sc iv}~}

\newcommand{\oii}{[O~{\sc ii}]~}

\newcommand{\hi}{H~{\sc i}~}

\newcommand{\mgiinsp}{Mg~{\sc ii}}

\newcommand{\mgii}{Mg~{\sc ii}~}

\newcommand{\angmath}{\textrm{\AA}}

\newcommand{\mmgii}{{\rm Mg \; \mbox{\tiny II}}}

\newcommand{\kms}{km s$^{-1}$}

%% file: mle_params.tex
\newcommand{\ewstar}{0.824}

\newcommand{\ealtvar}{0.090}
\newcommand{\wmineff}{0.337}
\newcommand{\wminefferr}{0.090}
\newcommand{\enstar}{1.827}
\newcommand{\enstarerr}{0.059}

\newcommand{\ewstarlow}{0.935}

\newcommand{\ealtvarlow}{0.150}
\newcommand{\wminefflow}{0.393}
\newcommand{\wminefferrlow}{0.150}

\newcommand{\ewstarmid}{0.766}

\newcommand{\ealtvarmid}{0.152}
\newcommand{\wmineffmid}{0.278}
\newcommand{\wminefferrmid}{0.152}

\newcommand{\ewstarhigh}{0.700}

\newcommand{\ealtvarhigh}{0.180}
\newcommand{\wmineffhigh}{0.269}
\newcommand{\wminefferrhigh}{0.180}

\newcommand{\pbeta}{0.133}
\newcommand{\pvar}{0.477}
\newcommand{\pnstar}{1.113}
\newcommand{\pnstarerr}{0.759}

\newcommand{\pebeta}{0.002}
\newcommand{\pewstar}{0.824}

\newcommand{\pealtvarone}{0.090}

\newcommand{\pealtvarfour}{0.487}
\newcommand{\penstar}{2.211}
\newcommand{\penstarerr}{1.533}

\newcommand{\zewstar}{1.166}
\newcommand{\zedelta}{-0.229}

\newcommand{\zealtvarone}{0.359}
\newcommand{\zealtvartwo}{-0.940}

\newcommand{\zealtvarfour}{0.220}
\newcommand{\zenstar}{1.814}
\newcommand{\zenstarerr}{0.057}

%% file: qsoList.tex
\begin{deluxetable}{l c c c c}
\tablecaption{FIRE \mgii Survey Sightlines}
\tablehead{ \colhead{Object} & \colhead{$z_{QSO}$} & \colhead{$\Delta z$} & \colhead{$t_{exp}$ } & \colhead{SNR\tablenotemark{a}} \\ \colhead{} & \colhead{} & \colhead{} & \colhead{$(s)$} & \colhead{} }
\startdata
Q0000-26 & 4.10 & 1.95-3.83 & 1226 & 20.7
 \\

BR0004-6224 & 4.51 & 1.95-4.51 & 1764 & 8.2
 \\

BR0016-3544 & 4.15 & 1.95-3.83 & 2409 & 14.0
 \\

SDSS0106+0048 & 4.45 & 1.95-4.45 & 3635 & 18.9
 \\

SDSS0113-0935 & 3.67 & 1.95-3.67 & 1944 & 12.8
 \\

SDSS0127-0045 & 4.08 & 1.95-3.83 & 3635 & 22.5
 \\

SDSS0140-0839 & 3.71 & 1.95-3.71 & 1226 & 18.2
 \\

SDSS0203+0012 & 5.85 & 1.98-5.40 & 3635 & 4.0
 \\

SDSS0244-0816 & 4.07 & 1.95-3.83 & 1944 & 12.9
 \\

BR0305-4957 & 4.78 & 1.95-4.78 & 2409 & 28.2
 \\

BR0322-2928 & 4.62 & 1.95-4.62 & 2409 & 21.1
 \\

BR0331-1622 & 4.32 & 1.95-4.32 & 1944 & 15.1
 \\

SDSS0331-0741 & 4.74 & 1.95-4.74 & 2177 & 6.2
 \\

SDSS0332-0654 & 3.69 & 1.95-3.69 & 2409 & 5.6
 \\

SDSS0344-0653 & 3.96 & 1.95-3.83 & 3022 & 6.6
 \\

BR0353-3820 & 4.58 & 1.95-4.58 & 1200 & 26.7
 \\

BR0418-5723 & 4.37 & 1.95-4.37 & 4200 & 8.5
 \\

SDSS0818+1722 & 5.90 & 2.00-5.40 & 9000 & 10.2
 \\

SDSS0836+0054 & 5.82 & 1.96-5.40 & 10187 & 15.8
 \\

SDSS0842+0637 & 3.66 & 1.95-3.66 & 2409 & 9.1
 \\

SDSS0935+0022 & 5.82 & 1.96-5.40 & 1817 & 11.1
 \\

SDSS0949+0335 & 4.05 & 1.95-3.83 & 1817 & 13.6
 \\

SDSS1020+0922 & 3.64 & 1.95-3.64 & 2409 & 15.2
 \\

SDSS1030+0524 & 6.28 & 2.16-6.28 & 14400 & 5.0
 \\

SDSS1037+0704 & 4.10 & 1.95-3.83 & 2726 & 8.1
 \\

SDSS1110+0244 & 4.12 & 1.95-3.83 & 2409 & 18.6
 \\

SDSS1135+0842 & 3.83 & 1.95-3.83 & 2409 & 17.7
 \\

SDSS1249-0159 & 3.64 & 1.95-3.64 & 1817 & 18.2
 \\

SDSS1305+0521 & 4.09 & 1.95-3.83 & 1363 & 8.8
 \\

SDSS1306+0356 & 5.99 & 2.04-5.99 & 15682 & 6.4
 \\

ULAS1319+0950 & 6.13 & 2.10-6.13 & 19275 & 5.0
 \\

SDSS1402+0146 & 4.16 & 1.95-3.83 & 1902 & 15.0
 \\

SDSS1408+0205 & 4.01 & 1.95-3.83 & 2409 & 9.9
 \\

SDSS1411+1217 & 5.93 & 2.01-5.93 & 3600 & 8.6
 \\

Q1422+2309 & 3.65 & 1.95-3.65 & 1226 & 47.2
 \\

SDSS1433+0227 & 4.72 & 1.95-4.72 & 2409 & 13.4
 \\

CFQS1509-1749 & 6.12 & 2.10-6.12 & 9900 & 9.7
 \\

SDSS1538+0855 & 3.55 & 1.95-3.55 & 1363 & 24.2
 \\

SDSS1616+0501 & 4.87 & 1.95-4.87 & 1800 & 5.2
 \\

SDSS1620+0020 & 4.09 & 1.95-3.83 & 972 & 7.0
 \\

SDSS1621-0042 & 3.70 & 1.95-3.70 & 1204 & 26.1
 \\

SDSS2147-0838 & 4.59 & 1.95-4.59 & 2409 & 13.8
 \\

SDSS2225-0014 & 4.89 & 1.95-4.89 & 7200 & 11.5
 \\

SDSS2228-0757 & 5.14 & 1.95-5.14 & 3600 & 4.9
 \\

SDSS2310+1855 & 6.04 & 2.06-6.04 & 14400 & 17.5
 \\

BR2346-3729 & 4.21 & 1.95-3.83 & 2409 & 11.0

\enddata                     
\tablenotetext{a}{Median signal-to-noise ratio per pixel across \mgii pathlength.}
\label{tab:qsoList}

\end{deluxetable}

%% file: dlist.tex
\begin{deluxetable*}{ c l c c c c c c}
\tablewidth{0pc}\tablecaption{Summary of Absorption Properties for the FIRE \mgii Sample}\tablehead{ \colhead{Index \#} & \colhead{Sightline} & \colhead{$z$} & \colhead{$W_{r}(2796)$} & \colhead{$\sigma(2796)$} & \colhead{$W_r(2803)$} & \colhead{$\sigma(2803)$} & \colhead{$\Delta v$} \\ \colhead{} & \colhead{} & \colhead{} & \colhead{(\AA)} & \colhead{(\AA)} & \colhead{(\AA)} & \colhead{(\AA)} & \colhead{(\kms)} }\startdata

1~ & Q0000-26 & 2.184 & 0.162 & 0.031 & 0.112 & 0.026 & 166.0 \\
2~ & Q0000-26 & 3.390 & 1.356 & 0.016 & 1.145 & 0.016 & 259.7 \\
3~ & BR0004-6224 & 2.663 & 0.260 & 0.045 & 0.140 & 0.036 & 58.0 \\
4~ & BR0004-6224 & 2.908 & 0.596 & 0.047 & 0.183 & 0.028 & 83.3 \\
5~ & BR0004-6224 & 2.959 & 0.569 & 0.063 & 0.669 & 0.047 & 108.4 \\
6~ & BR0004-6224 & 3.203 & 0.558 & 0.029 & 0.548 & 0.026 & 86.8 \\
7~ & BR0004-6224 & 3.694 & 0.236 & 0.042 & 0.234 & 0.019 & 79.6 \\
8~ & BR0004-6224 & 3.776 & 1.045 & 0.046 & 1.049 & 0.043 & 169.7 \\
9~ & BR0016-3544 & 2.783 & 0.517 & 0.027 & 0.305 & 0.026 & 101.2 \\
10~ & BR0016-3544 & 2.819 & 4.028 & 0.050 & 3.639 & 0.053 & 605.5 \\
11~ & BR0016-3544 & 2.949 & 0.157 & 0.026 & 0.144 & 0.035 & 79.6 \\
12~ & BR0016-3544 & 3.757 & 1.559 & 0.041 & 1.430 & 0.050 & 551.4 \\
13~ & SDSS0106+0048 & 3.729 & 0.842 & 0.016 & 0.673 & 0.015 & 194.9 \\
14~ & SDSS0113-0935 & 2.825 & 0.194 & 0.029 & 0.110 & 0.029 & 94.0 \\
15~ & SDSS0113-0935 & 3.544 & 0.228 & 0.037 & 0.186 & 0.035 & 133.6 \\
16~ & SDSS0113-0935 & 3.617 & 0.563 & 0.024 & 0.344 & 0.020 & 155.3 \\
17~ & SDSS0127-0045 & 2.588 & 1.602 & 0.025 & 1.164 & 0.025 & 457.8 \\
18~ & SDSS0127-0045 & 2.945 & 2.253 & 0.038 & 1.583 & 0.037 & 529.7 \\
19~ & SDSS0127-0045 & 3.168 & 0.309 & 0.024 & 0.138 & 0.016 & 198.5 \\
20~ & SDSS0127-0045 & 3.728 & 0.824 & 0.012 & 0.745 & 0.012 & 176.9 \\
21~ & SDSS0140-0839 & 2.241 & 0.405 & 0.031 & 0.686 & 0.043 & 144.4 \\
22\tablenotemark{a} & SDSS0140-0839 & 3.081 & 0.558 & 0.018 & 0.410 & 0.027 & 126.4 \\
23~ & SDSS0140-0839 & 3.212 & 0.081 & 0.014 & 0.092 & 0.014 & 126.4 \\
24\tablenotemark{b} & SDSS0140-0839 & 3.697 & 0.604 & 0.014 & 0.308 & 0.009 & 173.3 \\
25~ & SDSS0203+0012 & 3.711 & 0.374 & 0.038 & 0.250 & 0.065 & 122.8 \\
26\tablenotemark{a} & SDSS0203+0012 & 4.313 & 0.849 & 0.093 & 0.824 & 0.080 & 198.5 \\
27~ & SDSS0203+0012 & 4.482 & 0.670 & 0.183 & 0.623 & 0.024 & 126.5 \\
28~ & SDSS0203+0012 & 4.978 & 0.886 & 0.039 & 0.791 & 0.056 & 151.6 \\
29~ & BR0305-4957 & 2.502 & 0.331 & 0.024 & 0.169 & 0.022 & 137.2 \\
30~ & BR0305-4957 & 2.629 & 1.113 & 0.018 & 0.959 & 0.023 & 220.1 \\
31~ & BR0305-4957 & 3.354 & 0.564 & 0.013 & 0.412 & 0.012 & 144.5 \\
32~ & BR0305-4957 & 3.591 & 1.373 & 0.017 & 1.207 & 0.013 & 306.5 \\
33~ & BR0305-4957 & 4.466 & 1.792 & 0.017 & 1.478 & 0.029 & 302.9 \\
34~ & BR0322-2928 & 2.229 & 0.618 & 0.020 & 0.510 & 0.021 & 112.1 \\
35~ & BR0331-1622 & 2.295 & 1.836 & 0.067 & 1.714 & 0.056 & 403.8 \\
36~ & BR0331-1622 & 2.593 & 0.223 & 0.019 & 0.185 & 0.019 & 76.0 \\
37~ & BR0331-1622 & 2.927 & 1.382 & 0.039 & 1.098 & 0.045 & 346.1 \\
38~ & BR0331-1622 & 3.557 & 0.707 & 0.033 & 0.582 & 0.033 & 130.0 \\
39\tablenotemark{a} & SDSS0332-0654 & 3.061 & 0.883 & 0.084 & 0.608 & 0.059 & 162.4 \\
40~ & BR0353-3820 & 1.987 & 3.131 & 0.030 & 2.717 & 0.026 & 508.2 \\
41~ & BR0353-3820 & 2.696 & 0.381 & 0.014 & 0.232 & 0.014 & 187.7 \\
42~ & BR0353-3820 & 2.754 & 4.599 & 0.016 & 4.325 & 0.019 & 616.3 \\
43~ & BR0418-5723 & 2.030 & 1.449 & 0.072 & 1.009 & 0.080 & 209.3 \\
44\tablenotemark{a} & BR0418-5723 & 2.978 & 1.850 & 0.072 & 2.136 & 0.099 & 263.3 \\
45~ & SDSS0818+1722 & 3.563 & 0.640 & 0.072 & 0.427 & 0.029 & 151.7 \\
46~ & SDSS0818+1722 & 4.431 & 0.457 & 0.052 & 0.138 & 0.010 & 108.4 \\
47~ & SDSS0818+1722 & 5.065 & 0.841 & 0.061 & 0.533 & 0.046 & 151.7 \\
48~ & SDSS0836+0054 & 2.299 & 0.455 & 0.022 & 0.300 & 0.021 & 130.0 \\
49~ & SDSS0836+0054 & 3.744 & 2.607 & 0.024 & 1.992 & 0.031 & 443.4 \\
50~ & SDSS0949+0335 & 2.289 & 2.852 & 0.062 & 2.408 & 0.054 & 421.8 \\
51~ & SDSS0949+0335 & 3.310 & 2.033 & 0.039 & 1.665 & 0.033 & 339.0 \\
52~ & SDSS1020+0922 & 2.046 & 0.406 & 0.045 & 0.288 & 0.047 & 112.0 \\
53~ & SDSS1020+0922 & 2.593 & 0.464 & 0.026 & 0.499 & 0.022 & 144.5 \\
54~ & SDSS1020+0922 & 2.749 & 0.652 & 0.023 & 0.518 & 0.024 & 151.7 \\
55~ & SDSS1020+0922 & 3.479 & 0.118 & 0.016 & 0.085 & 0.019 & 104.8 \\
56~ & SDSS1030+0524 & 2.188 & 0.317 & 0.017 & 0.291 & 0.017 & 133.7 \\
57\tablenotemark{c} & SDSS1030+0524 & 2.780 & 2.617 & 0.069 & 1.855 & 0.086 & 583.9 \\
58~ & SDSS1030+0524 & 4.583 & 1.857 & 0.031 & 2.139 & 0.127 & 367.7 \\
59~ & SDSS1030+0524 & 4.948 & 0.447 & 0.017 & 0.278 & 0.019 & 162.4 \\
60~ & SDSS1030+0524 & 5.130 & 0.138 & 0.013 & 0.089 & 0.023 & 76.0
\enddata
\label{tab:dlist}
\end{deluxetable*}

%% file: dlistb.tex
\begin{deluxetable*}{ c l c c c c c c}
\tablewidth{0pc}\tablecaption{Summary of Absorption Properties for the FIRE \mgii Sample (\emph{Continued})}\tablehead{ \colhead{Index \#} & \colhead{Sightline} & \colhead{$z$} & \colhead{$W_{r}(2796)$} & \colhead{$\sigma(2796)$} & \colhead{$W_r(2803)$} & \colhead{$\sigma(2803)$} & \colhead{$\Delta v$} \\ \colhead{} & \colhead{} & \colhead{} & \colhead{(\AA)} & \colhead{(\AA)} & \colhead{(\AA)} & \colhead{(\AA)} & \colhead{(\kms)} }\startdata

61~ & SDSS1110+0244 & 2.119 & 3.041 & 0.041 & 2.884 & 0.042 & 454.2 \\
62~ & SDSS1110+0244 & 2.223 & 0.205 & 0.024 & 0.121 & 0.029 & 166.0 \\
63~ & SDSS1305+0521 & 2.302 & 1.993 & 0.095 & 1.533 & 0.095 & 234.5 \\
64~ & SDSS1305+0521 & 2.753 & 0.378 & 0.040 & 0.319 & 0.037 & 140.9 \\
65~ & SDSS1305+0521 & 3.235 & 0.328 & 0.025 & 0.130 & 0.025 & 122.8 \\
66~ & SDSS1305+0521 & 3.680 & 1.781 & 0.068 & 1.583 & 0.039 & 353.4 \\
67~ & SDSS1306+0356 & 2.533 & 3.307 & 0.101 & 3.019 & 0.088 & 594.6 \\
68~ & SDSS1306+0356 & 3.490 & 0.648 & 0.031 & 0.526 & 0.044 & 169.7 \\
69~ & SDSS1306+0356 & 4.615 & 0.983 & 0.078 & 0.724 & 0.038 & 187.7 \\
70~ & SDSS1306+0356 & 4.865 & 2.798 & 0.044 & 3.049 & 0.087 & 403.7 \\
71~ & SDSS1306+0356 & 4.882 & 1.941 & 0.079 & 2.276 & 0.040 & 248.8 \\
72~ & ULAS1319+0950 & 4.569 & 0.406 & 0.062 & 0.177 & 0.029 & 101.2 \\
73~ & SDSS1402+0146 & 3.277 & 1.075 & 0.018 & 1.034 & 0.028 & 234.4 \\
74\tablenotemark{c} & SDSS1402+0146 & 3.454 & 0.341 & 0.016 & 0.112 & 0.018 & 173.3 \\
75~ & SDSS1408+0205 & 1.982 & 2.174 & 0.056 & 1.769 & 0.054 & 252.5 \\
76~ & SDSS1408+0205 & 1.991 & 0.914 & 0.041 & 0.555 & 0.042 & 126.4 \\
77~ & SDSS1408+0205 & 2.462 & 1.385 & 0.040 & 1.029 & 0.035 & 209.3 \\
78~ & SDSS1411+1217 & 2.237 & 0.627 & 0.041 & 0.334 & 0.045 & 148.0 \\
79~ & SDSS1411+1217 & 3.477 & 0.343 & 0.016 & 0.179 & 0.022 & 83.2 \\
80~ & SDSS1411+1217 & 4.929 & 0.644 & 0.023 & 0.488 & 0.018 & 317.2 \\
81~ & SDSS1411+1217 & 5.055 & 0.207 & 0.013 & 0.092 & 0.015 & 90.4 \\
82~ & SDSS1411+1217 & 5.250 & 0.330 & 0.013 & 0.190 & 0.011 & 130.1 \\
83~ & SDSS1411+1217 & 5.332 & 0.197 & 0.013 & 0.241 & 0.011 & 97.6 \\
84\tablenotemark{c} & Q1422+2309 & 3.540 & 0.169 & 0.011 & 0.115 & 0.008 & 130.0 \\
85~ & SDSS1433+0227 & 2.772 & 0.735 & 0.018 & 0.601 & 0.024 & 166.0 \\
86\tablenotemark{a} & CFQS1509-1749 & 3.128 & 0.858 & 0.093 & 0.773 & 0.043 & 220.1 \\
87~ & CFQS1509-1749 & 3.266 & 0.896 & 0.021 & 0.711 & 0.023 & 194.9 \\
88~ & CFQS1509-1749 & 3.392 & 5.585 & 0.071 & 5.082 & 0.050 & 709.8 \\
89~ & SDSS1538+0855 & 2.638 & 0.278 & 0.027 & 0.206 & 0.028 & 180.4 \\
90~ & SDSS1538+0855 & 3.498 & 0.151 & 0.011 & 0.122 & 0.014 & 173.2 \\
91~ & SDSS1616+0501 & 2.741 & 1.510 & 0.044 & 0.923 & 0.051 & 169.7 \\
92~ & SDSS1616+0501 & 3.275 & 0.600 & 0.036 & 0.494 & 0.110 & 104.8 \\
93~ & SDSS1616+0501 & 3.396 & 0.960 & 0.036 & 0.631 & 0.113 & 126.4 \\
94~ & SDSS1616+0501 & 3.450 & 0.606 & 0.033 & 0.557 & 0.053 & 115.6 \\
95~ & SDSS1616+0501 & 3.733 & 2.252 & 0.189 & 1.421 & 0.068 & 263.3 \\
96~ & SDSS1620+0020 & 2.910 & 1.130 & 0.058 & 1.063 & 0.058 & 263.3 \\
97~ & SDSS1620+0020 & 3.273 & 0.965 & 0.043 & 0.635 & 0.052 & 202.1 \\
98~ & SDSS1620+0020 & 3.620 & 1.357 & 0.065 & 1.091 & 0.042 & 320.9 \\
99~ & SDSS1620+0020 & 3.752 & 1.656 & 0.065 & 1.550 & 0.095 & 374.9 \\
100~ & SDSS1621-0042 & 2.678 & 0.176 & 0.017 & 0.135 & 0.016 & 97.6 \\
101\tablenotemark{a} & SDSS1621-0042 & 3.106 & 0.974 & 0.011 & 1.011 & 0.012 & 112.0 \\
102~ & SDSS2147-0838 & 2.286 & 0.977 & 0.040 & 0.567 & 0.033 & 158.9 \\
103~ & SDSS2228-0757 & 3.175 & 0.304 & 0.037 & 0.243 & 0.031 & 50.8 \\
104~ & SDSS2310+1855 & 2.243 & 1.441 & 0.050 & 0.781 & 0.049 & 292.1 \\
105~ & SDSS2310+1855 & 2.351 & 0.807 & 0.042 & 0.492 & 0.035 & 212.9 \\
106~ & SDSS2310+1855 & 2.643 & 0.863 & 0.036 & 0.339 & 0.059 & 173.3 \\
107~ & SDSS2310+1855 & 3.300 & 0.665 & 0.039 & 0.457 & 0.034 & 212.9 \\
108~ & BR2346-3729 & 2.830 & 1.633 & 0.049 & 1.421 & 0.037 & 238.1 \\
109~ & BR2346-3729 & 2.923 & 0.557 & 0.030 & 0.636 & 0.034 & 162.5 \\
110~ & BR2346-3729 & 3.619 & 0.412 & 0.031 & 0.240 & 0.019 & 151.6 \\
111~ & BR2346-3729 & 3.692 & 0.385 & 0.016 & 0.413 & 0.046 & 133.6
\enddata
\tablenotetext{a}{Poor telluric region}
\tablenotetext{b}{Proximate system}
\tablenotetext{c}{Missed by automated search algorithm}
\nonumber
\label{}
\end{deluxetable*}

%% file: dndws_var.tex
\begin{deluxetable}{c c c c c}
\tablecaption{\mgii Equivalent Width Distribution, \\ Full Sample and Redshift Cuts}
\tablehead{ \colhead{$\left<W_r\right>$} & \colhead{$\Delta W_r$} & \colhead{$\bar{C}$} & \colhead{Number} & \colhead{$d^2N/dzdW$} \\ \colhead{(\AA)} & \colhead{(\AA)} & \colhead{$(\%)$} & \colhead{} & \colhead{}}
\multicolumn{5}{c}{$z = 1.90-6.30$} \\
\hline
0.42 & 0.05-0.64 & 50.0 & 47 & 1.570$\pm$0.272 \\
0.94 & 0.64-1.23 & 81.2 & 28 & 0.594$\pm$0.119 \\
1.52 & 1.23-1.82 & 82.6 & 14 & 0.291$\pm$0.080 \\
2.11 & 1.82-2.41 & 82.6 & 9 & 0.187$\pm$0.064 \\
2.70 & 2.41-3.00 & 82.6 & 4 & 0.083$\pm$0.042 \\
4.34 & 3.00-5.68 & 82.6 & 6 & 0.027$\pm$0.011 \\ 
\hline

\multicolumn{5}{c}{$z = 1.95-2.98$} \\
\hline
0.43 & 0.05-0.64 & 44.3 & 22 & 1.712$\pm$0.412 \\
0.94 & 0.64-1.23 & 77.5 & 8 & 0.381$\pm$0.137 \\
1.53 & 1.23-1.82 & 79.2 & 7 & 0.323$\pm$0.125 \\
2.12 & 1.82-2.41 & 79.2 & 4 & 0.187$\pm$0.094 \\
2.71 & 2.41-3.00 & 79.2 & 2 & 0.093$\pm$0.066 \\
4.34 & 3.00-5.68 & 79.2 & 5 & 0.051$\pm$0.023 \\ 
\hline

\multicolumn{5}{c}{$z = 3.15-3.81$} \\
\hline
0.41 & 0.05-0.64 & 61.7 & 17 & 1.634$\pm$0.414 \\
0.94 & 0.64-1.23 & 91.1 & 11 & 0.723$\pm$0.222 \\
1.52 & 1.23-1.82 & 92.0 & 6 & 0.391$\pm$0.161 \\
2.11 & 1.82-2.41 & 92.0 & 2 & 0.130$\pm$0.092 \\
2.70 & 2.41-3.00 & 92.0 & 1 & 0.065$\pm$0.065 \\
4.34 & 3.00-5.68 & 92.0 & 1 & 0.014$\pm$0.014 \\ 
\hline

\multicolumn{5}{c}{$z = 4.34-5.35$} \\
\hline
0.41 & 0.05-0.64 & 61.0 & 7 & 1.429$\pm$0.559 \\
0.94 & 0.64-1.23 & 90.7 & 5 & 0.694$\pm$0.313 \\
1.52 & 1.23-1.82 & 91.5 & 1 & 0.138$\pm$0.138 \\
2.11 & 1.82-2.41 & 91.5 & 2 & 0.275$\pm$0.195 \\
2.70 & 2.41-3.00 & 91.5 & 1 & 0.138$\pm$0.138 \\
4.34 & 3.00-5.68 & 91.5 & 0 & $<$0.030
\enddata

\label{tab:dndws_var}

\end{deluxetable}

%% file: wstars.tex
\begin{deluxetable}{c c c c c}
\tablecaption{Maximum-Likelihood Fit Parameters for \\ Exponential Parameterization of the $W_r$ Distribution}
\tablehead{ \colhead{$\left< z \right>$} & \colhead{$\Delta z$} & \colhead{$W^*$} & \colhead{$N^*$} \\ \colhead{} & \colhead{} & \colhead{(\AA)} & \colhead{} }
0.68\tablenotemark{a} & 0.366-0.871 & 0.585$\pm$0.024 & 1.216$\pm$0.124 \\
1.10\tablenotemark{a} & 0.871-1.311 & 0.741$\pm$0.032 & 1.171$\pm$0.083 \\
1.60\tablenotemark{a} & 1.311-2.269 & 0.804$\pm$0.034 & 1.267$\pm$0.092 \\
2.51 & 1.947-2.975 & 0.935$\pm$0.150 & 1.755$\pm$0.081 \\
3.46 & 3.150-3.805 & 0.766$\pm$0.152 & 1.952$\pm$0.105 \\
4.78 & 4.345-5.350 & 0.700$\pm$0.180 & 1.811$\pm$0.141 \\
3.24 & 1.947-6.207 & 0.824$\pm$0.090 & 1.827$\pm$0.059
\enddata
\tablenotetext{a}{Parameter fits from \citet{nestor2005}}\label{tab:wstars}

\end{deluxetable}

%% file: dndzs.tex
\begin{deluxetable}{c c c c c c}
\tablecaption{\mgii Absorption line density $dN/dz$}
\tablehead{ \colhead{$\left< z \right>$} & \colhead{$\Delta z$} & \colhead{$\bar{C}$} & \colhead{Number} & \colhead{$dN/dz$} & \colhead{$dN/dX$} \\ \colhead{} & \colhead{} & \colhead{$(\%)$} & \colhead{} & \colhead{} & \colhead{} }
\startdata

\multicolumn{6}{c}{$0.3 \angmath <W_r<0.6$ \angmath}\\
\hline
2.231 & 1.947-2.461 & 47.3 & 4 & 0.362$\pm$0.193 & 0.115$\pm$0.061 \\
2.727 & 2.461-2.975 & 69.1 & 8 & 0.486$\pm$0.176 & 0.141$\pm$0.051 \\
3.457 & 3.150-3.805 & 77.9 & 10 & 0.453$\pm$0.146 & 0.119$\pm$0.038 \\
4.786 & 4.345-5.350 & 76.9 & 4 & 0.387$\pm$0.195 & 0.088$\pm$0.045 \\ 
\hline
\multicolumn{6}{c}{$0.6 \angmath<W_r<1.0 \angmath$}\\
\hline
2.232 & 1.947-2.461 & 65.5 & 5 & 0.343$\pm$0.156 & 0.109$\pm$0.049 \\
2.723 & 2.461-2.975 & 84.6 & 3 & 0.149$\pm$0.087 & 0.043$\pm$0.025 \\
3.458 & 3.150-3.805 & 89.9 & 11 & 0.432$\pm$0.132 & 0.113$\pm$0.035 \\
4.779 & 4.345-5.350 & 89.6 & 5 & 0.415$\pm$0.187 & 0.095$\pm$0.043 \\ 
\hline
\multicolumn{6}{c}{$W_r>1.0 \angmath$}\\
\hline
2.231 & 1.947-2.461 & 70.4 & 8 & 0.509$\pm$0.185 & 0.161$\pm$0.059 \\
2.722 & 2.461-2.975 & 87.5 & 12 & 0.578$\pm$0.170 & 0.168$\pm$0.050 \\
3.459 & 3.150-3.805 & 92.0 & 12 & 0.461$\pm$0.135 & 0.121$\pm$0.035 \\
4.776 & 4.345-5.350 & 91.5 & 4 & 0.325$\pm$0.164 & 0.074$\pm$0.037
\enddata

\label{tab:dndzs}

\end{deluxetable}

%% file: betas.tex
\begin{deluxetable}{c c c c c}
\tablecaption{Maximum-Likelihood Estimates of the \\ Line Density Evolution $dN/dz=N^*(1+z)^\beta$}
\tablehead{ \colhead{$\left< W_r \right>$} & \colhead{$\Delta W_r$} & \colhead{$\Delta z$} & \colhead{$\beta$} & \colhead{$N^*$} \\ \colhead{(\AA)} & \colhead{(\AA)} & \colhead{} & \colhead{} & \colhead{} }\\
\startdata
1.17\tablenotemark{a} & 1.00-1.40 & 0.35-2.3 & $0.99^{+0.29}_{-0.22}$ & $0.51^{+0.09}_{-0.10}$ \\
1.58\tablenotemark{a} & 1.40-1.80 & 0.35-2.3 & $1.56^{+0.33}_{-0.31}$ & $0.020^{+0.05}_{-0.05}$ \\
1.63\tablenotemark{a} & 1.00+ & 0.35-2.3 & $1.40^{+0.16}_{-0.16}$ & $0.08^{+0.15}_{-0.05}$ \\
2.08\tablenotemark{a} & 1.40+ & 0.35-2.3 & $1.74^{+0.22}_{-0.22}$ & $0.036^{+0.06}_{-0.06}$ \\
2.52\tablenotemark{a} & 1.80+ & 0.35-2.3 & $1.92^{+0.30}_{-0.32}$ & $0.016^{+0.06}_{-0.03}$ \\
$\ewstar$ & Full Range & 1.9-6.3 & $\pbeta \pm \pvar$ & $\pnstar \pm \pnstarerr$ \\
0.45 & 0.30-0.60 & 1.9-6.3 & -0.104$\pm$0.937 & 0.482$\pm$0.644 \\
0.79 & 0.60-1.00 & 1.9-6.3 & 0.928$\pm$0.889 & 0.093$\pm$0.121 \\
1.82 & 1.00+ & 1.9-6.3 & -0.746$\pm$0.857 & 1.301$\pm$1.555
\enddata\tablenotetext{a}{Parameter fits from \citet{prochtersdss2006}, with corresponding upper and lower $95\%$ confidence intervals.  This survey's results include $1\sigma$ errors.}\label{tab:betas}

\end{deluxetable}